\pdfoutput=1   
\documentclass[usenatbib,usegraphicx,fleqn]{mn2e}
\usepackage{bm,color}
\usepackage{charter}   
\usepackage{graphicx, amsmath, amssymb, wasysym, natbib}
\newcommand{\msun}{\ensuremath{{\rm M}_{\odot}}} 
 
\newcommand{\rsun}{\ensuremath{{\rm R}_{\odot}}}
\newcommand{\mstar}{\ensuremath{{\rm M}_{*}}}
\newcommand{\rstar}{\ensuremath{{\rm R}_{*}}}

\newcommand{\mdot}{\ensuremath{\dot{\text{M}}}} 
\newcommand{\ldot}{\ensuremath{\dot{\text{L}}}} 
\newcommand{\bvec}{\ensuremath{\mathbf{B}}} 
\newcommand{\trho}{\ensuremath{\rho}}
\newcommand{\trad}{\ensuremath{\text{r}}}

\newcommand{\ttim}{\ensuremath{\text{t}}}
\newcommand{\tvel}{\ensuremath{\text{v}}}
\newcommand{\rhov}{\ensuremath{\rho \left| {\mathbf{v}_p} \right|}}
\newcommand{\vvec}{\ensuremath{\mathbf{v}}}
\newcommand{\gvec}{\ensuremath{\mathbf{g}}}
\newcommand{\rvec}{\ensuremath{\mathbf{r}}}
\newcommand{\tensor}{\ensuremath{\mathcal{T}}}
\newcommand{\ts}{\ensuremath{s}}
\font\tenbg=cmmib10 at 10pt

\def \rvecphi{{\hbox{\tenbg\char'036}}}
\def \rvectheta{{\hbox{\tenbg\char'022}}}

\def\lesssim{\mathrel{\hbox{\rlap{\hbox{\lower4pt\hbox{$\sim$}}}\hbox{$<$}}}}
\def\gtrsim{\mathrel{\hbox{\rlap{\hbox{\lower4pt\hbox{$\sim$}}}\hbox{$>$}}}}

\font\tenbg=cmmib10 at 10pt

\def \rvecphi{{\hbox{\tenbg\char'036}}}
\def \rvectheta{{\hbox{\tenbg\char'022}}}

\def\lesssim{\mathrel{\hbox{\rlap{\hbox{\lower4pt\hbox{$\sim$}}}\hbox{$<$}}}}
\def\gtrsim{\mathrel{\hbox{\rlap{\hbox{\lower4pt\hbox{$\sim$}}}\hbox{$>$}}}}

\title[Magnetic launching and collimation of jets in 2.5D MHD simulations]{Magnetic launching and collimation of jets from the disk-magnetosphere boundary: 2.5D MHD simulations}
\author[Lii, Romanova \& Lovelace]{Patrick Lii\thanks{E-mail of corresponding author: pslii@astro.cornell.edu}, Marina Romanova, Richard Lovelace \\
Department of Astronomy, Cornell University, Ithaca, NY 14853}

\begin{document}
\date{\today}

\pagerange{\pageref{firstpage}--\pageref{lastpage}} \pubyear{2011}

\maketitle

\label{firstpage}

\begin{abstract}
We use axisymmetric magnetohydrodynamic (MHD) simulations to investigate the launching and collimation of jets emerging from the disk-magnetosphere boundary of accreting magnetized stars. Our analysis shows that the matter flows into the jet from the inner edge of the accretion disk. It is magnetically accelerated along field lines extending up from the disk and simultaneously collimated by the magnetic pinch force. In the reference run which we use for analysis, the matter in the jet crosses the Alfv\'en surface a few $R_*$ above the disk and the fast magnetosonic surface  $\sim$13 $R_*$ above the disk. At larger distances, the magnetic pressure is a few times smaller than the total matter pressure but the magnetic force continues to accelerate and collimate the jet. In steady state, we observe a matter ejection-to-accretion ratio of $\sim$0.2. Across different simulation runs, we measure a range of half-opening angles between $\Theta \approx 4^\circ$ and $20^\circ$ at the top of the simulation region, depending on the degree of magnetization in the outflow. We consider the case of stars undergoing epochs of high accretion (such as EXors, FUORs, and CTTSs) where the stellar magnetosphere is strongly compressed by the incoming accretion disk. For a typical EXor (mass 0.8 \msun, radius 2 \rsun) accreting at $\sim10^{-5}$ \msun/yr, we measure poloidal velocities in the jet ranging from 30 km/s on the outer edge of the jet to more than 260 km/s on the inner edge. In general, the models can be applied to a variety of magnetized stars---white dwarfs, neutron stars, and brown dwarfs---which exhibit periods of high accretion.
\end{abstract}

\section{Introduction} \label{sec_intro}
Young, accreting stars produce strongly collimated high-speed outflows (jets) which are a key mechanism in transporting mass, energy and angular momentum out of the disk and facilitating accretion onto the young protostellar core. The formation of a collimated jet requires tandem processes: the matter must be accelerated out of the disk and undergo collimation into a jet. A number of theoretical models have been proposed to explain the mechanism which launches the matter from the disk \citep[see review by][]{ferreira2006}. The first type is the magneto-centrifugal mechanism in which the outflow is launched from the disk as an extended wind \citep[][]{blandford1982, konigl2000}. A second class of models involves outflows driven from the the disk-magnetosphere boundary, either as an X-wind type outflow \citep{najita1994, shu1994a, cai2008} or as a magnetically launched conical wind \citep[][hereafter referred to as R09]{romanova2009}. The matter may also be magnetically launched from the inner disk \citep{lovelace1991} or driven as a stellar wind \citep{matt2008}. Observations of stars with strong outflows show that the typical outflow velocities are of the order of the Keplerian velocity of the inner disk region, favoring the models where the outflow originates from the disk-magnetosphere boundary or the inner disk. As the matter flows out of the disk, helical magnetic field lines frozen into the rotating outflow can self-collimate the matter into a jet \citep*{lovelace1987}. The outflow may also become collimated by interaction with an external poloidal field threading the disk \citep{matt2003, fendt2009}. Yet another possibility is that the matter becomes collimated by pressure from an external medium \citep*{lovelace1991, frank1996}.

Spectral measurements show that a significant number of Classical T Tauri Stars (CTTSs) exhibit signs of outflows from their disks \citep{edwards2003, edwards2006, gomezdecastro2010} with early observations indicating a correlation between the outflow and mass accretion rates of CTTSs \citep{cabrit1990, hartigan1995}. Even with improvements to telescopes, direct imaging of the inner disk region is still hampered by the insufficient resolution of ground- and space-based observatories \citep[see review by][]{ray2007} and observations have only recently shown that the jets become collimated at distances {\it less than} about 10 AU from the star \citep{hartigan2004, coffey2008}. In lieu of direct imaging of the inner disk, astronomers have relied on numerical simulations to study the structure and dynamics of the disk-magnetosphere boundary.

\paragraph*{Simulations} Early simulations performed by \citet*{hayashi1996} and \citet*{miller1997} achieved single-episode outflows from the disk-magnetosphere boundary lasting a few dynamical timescales. As simulations grew more sophisticated, longer runs performed by \citet{goodson1997, goodson1999}, \citet{hirose1997}, \citet{matt2002}, and \citet{kuker2003} showed several episodes of field line inflation and outflows. However, none of these early simulations produced robust outflows which lasted long enough to establish the outflow behavior and dynamics. Long lasting outflows have been achieved by treating the disk as a {\it boundary condition} and launching matter into the corona \citep[e.g.][]{romanova1997, ouyed1997, ustyugova1999, krasnopolsky1999, fendt2000, matsakos2008, fendt2009, staff2010}. However, In order to understand the launching and collimation mechanisms of the outflows, the simulations must include a realistic, low-temperature accretion disk and solve the full magnetohydrodynamic (MHD) equations in both the disk and coronal space. Recently, there has been much work in this direction with the modeling of outflows launched from realistic disks threaded by large scale magnetic fields \citep{casse2002, casse2004, zanni2007, murphy2010}.

Our group first obtained long lasting outflows from realistic accretion disks for the case of rapidly rotating stars in the ``propeller'' regime \citep{romanova2005, ustyugova2006}. These simulations showed a two-component outflow in which most of the matter is carried away through a conical-shaped wind while most of the energy and angular momentum flows into a low-density, high-velocity axial jet. This axial jet is magnetically-dominated and well collimated by the toroidal magnetic field. More recently, we observed single-component, long lasting conical outflows in the more general case of slowly rotating stars (R09). In R09, the conical outflows are driven from the disk-magnetosphere boundary by magnetic pressure and unlike X-winds, these outflows do not require equality of the magnetospheric and corotation radii and can originate even from slowly rotating stars. However, the conical winds showed only weak collimation within the simulation region.

In this work, we present axisymmetric MHD simulations of long-lasting, collimated jets launched from the disk-magnetosphere boundary of a magnetized star with an aligned dipole. Building upon the previous work on conical winds by R09, we simulate an accretion disk with a higher accretion rate in a larger simulation region and observe the emergence of a magnetically driven, magnetically collimated jet. In \S \ref{sec_methods} we describe the numerical methods and initial and boundary conditions used in the simulations. Then, in \S \ref{sec_overview}, we give an overview of the simulations and describe the reference run which we use for analysis of the launching and collimation mechanisms. In \S \ref{sec_analysis} we describe the fluxes, velocities and forces in the jet and investigate the jet launching and collimation mechanisms. Lastly, in \S \ref{sec_discussion} we discuss the relation of this work to previous results as well as the application of our simulations to several types of rapidly accreting young stars.

\section{Numerical setup} \label{sec_methods}
Here, we briefly describe the main aspects of our numerical model. The numerical model used here is identical to the setup used in R09 and utilizes an axisymmetric Godunov-type code based on the equations of viscous, resistive MHD. In the inertial reference frame, these equations are
\begin{align}
\frac{\partial\rho}{\partial t} + \nabla \cdot (\rho \vvec) & = 0, \\
\frac{\partial(\rho\vvec)}{\partial t} + \nabla \cdot \tensor & = \rho \gvec, \label{eqn_mhdmom} \\
\frac{\partial \bvec}{\partial t} - \nabla \times (\vvec \times \bvec) + \nabla \times (\eta_t\nabla \times \bvec) & = 0, \label{eqn_mhd} \\ 
\frac{\partial(\rho S)}{\partial t} + \nabla \cdot (\rho S \vvec) & = Q. 
\end{align}
In this set of equations, $\rho$ is the density, \vvec\ is the flow velocity, \tensor\ is the momentum flux-density tensor, $\gvec\ = -(G M / r^2)\hat{r}$  is the gravitational acceleration due to the star, \bvec\ is the magnetic field, $\eta_t$ is the magnetic diffusivity coefficient (see Appendix \ref{subsec_viscosity}), $S$ is the specific entropy, and $Q$ is the rate of change of entropy per unit volume. We treat the plasma as a monatomic ideal gas such that $S$ = $C_V \ln(P/\rho^\gamma)$ where $P$ is the gas pressure and the adiabatic index $\gamma$ = 5/3. 

\paragraph*{Grid} The code uses a spherical coordinate system ($r, \theta, \phi$) where $\theta$ is the colatitude and $\phi$ is the azimuth angle. Axisymmetry imposes the additional condition $\partial/\partial\phi$ = 0 on the equations of MHD\footnote{The equations of axisymmetric MHD in spherical coordinates can be found in \citet{ustyugova2006}.}. Since we use a Godunov scheme, all numerically calculated variables are cell centered on the grid, except for the magnetic vector potential which is calculated on the nodes. The simulations are performed in the region $R_* <\ r <\ R_{\rm out}$, $0 \le \theta \le \pi/2$ and reflected across the axisymmetry axis for plotting (the ``$z$-axis'' in cylindrical coordinates). The grid is uniform in the $\theta$ direction and the size steps in the radial direction are chosen such that the poloidal-plane cells are curvilinear rectangles with approximately equal lengths on each side. This choice results in high spatial resolution near the star where the disk-magnetosphere interaction takes place while also permitting a large simulation region. In the simulations presented in this paper, we choose the number of grids in the $\theta$ and $r$ directions to be $N_\theta$ = 50 and $N_r$ = 120, respectively. This corresponds to $d\theta = \pi/N_\theta = 1.8^\circ$ and a total region radius of about 42 \rstar. The smallest grid size at the surface of the star is $(\pi/100) \rstar$ on a side.

\subsection{Dimensionless variables}
Within the numerical code, the equations of MHD are reparametrized with normalized variables (e.g. $\tilde{\rho} = \rho/\rho_0$, $\tilde{B} = B/B_0$, $\tilde{v} = v/v_0$, etc.) and solved in a dimensionless form. To further simplify the equations, we also take $GM_*=1$ and ${\cal R}=1$. The dimensionless equations permit us to apply the general results to a wide variety of accreting stars. The calculation of the reference units is given in Appendix \ref{appen_units} and sample reference values for EX Lupi (EXor), FU Orionis (FUOR) and CTTS class stars are shown in Tab. \ref{tbl_refval}. In order to convert the dimensionless units into real values, multiply the dimensionless value by the corresponding reference value in Tab. \ref{tbl_refval}. For example, to calculate \ttim\ = 1000 for a typical CTTS, multiply by the reference time ($t_{0, {\rm CTTS}} $ = 0.366 days) to get t = 366 days.

For the remainder of the paper, all values and variables are given in terms of the normalized units (with tildes implicit) except where explicitly assigned physical units.

\subsection{Numerical method}
To numerically integrate the MHD equations, we split the physical processes into four blocks: (1) an ``ideal MHD" block in which we calculate the dynamics of the plasma and magnetic field without dissipative processes; two ``diffusion" blocks (2) and (3) in which we calculate the diffusion of the poloidal and azimuthal components of the magnetic field for frozen values of the plasma velocity and thermodynamic parameters (density and pressure); and a ``viscosity" block (4) in which we calculate the viscous dissipation due to the $r \phi$ and $\theta \phi$ components of the viscous stress tensor\footnote{See Appendix \ref{subsec_viscosity} for an in-depth discussion of the treatment of the turbulent viscosity and diffusivity in the simulations.}. Integration of the equations in time is performed with a two-step Runge-Kutta method. To determine the fluxes between the cells, we use an approximate solution of the Riemann problem analogous to the Roe solver described in \citet{brio1988}, except that we take the energy conservation equation in the entropy form. The dynamical variables are determined in the cells while the vector-potential of the magnetic field, $A_\phi$, is determined on the corner nodes. We guarantee the absence of magnetic charge by calculating $A_{\phi}$ at each time step and then using it to obtain the poloidal components of the magnetic field ($B_r$, $B_\theta$) in a divergence-free form \citep{toth2000}. This ensures that the divergence-free condition $\nabla \cdot B=0$ is always satisfied to within machine accuracy everywhere in the simulation region. We set a floor density of $\rho_{flr} = 2.5 \times 10^{-7}$ throughout the simulation region to prevent the density from vanishing near the axis. While the jet is being launched, this adds a small amount of matter to the grids nearest the axis.

For a complete description of the numerical method as well as standard tests of the code, see Appendices C and D in R09. 

\subsection{Initial and boundary conditions}
\paragraph*{Initial Conditions}
We take the stellar magnetic field to be an aligned dipole $\bvec_* = [3(\boldsymbol{\mu} \cdot \rvec) \rvec - \boldsymbol{\mu} r^2]/r^5$ where $\mu$ is the stellar dipole moment which is taken as a parameter in the simulations. Initially, the entire simulation region is filled with a non-rotating low-density, high-temperature isothermal plasma and there is no disk present. In spherical coordinates, the initial density and pressure distributions are
\begin{equation}
\rho=\rho_c\exp[GM/({\cal R} T_c r)], \quad P=P_c\exp[GM/({\cal R} T_c r)], \label{eqn_initial}
\end{equation}
where $T_c$ is the coronal temperature and $\rho_c$ is the coronal density at the external boundary, related by the ideal gas equation of state $P_c=\rho_c{\cal R} T_c$. The star initially corotates with the outer boundary, such that $\Omega_{*,i} = (G M_* / R_{\rm out}^3)^{1/2}$. Between \ttim\ = 0 and \ttim\ = 100 the star is gradually spun up, reaching a final value of $\Omega_{*, f} = [G M_* / (3 R_{*})^3]^{1/2} = 0.19$ corresponding to a disk corotation radius $R_{\rm cor} = 3$. Information about the stellar rotation propagates out at the Alfv$\acute{\rm e}$n speed along the stellar magnetic field lines into the low-density corona.

\paragraph*{External boundary} The external boundary $R_{\rm out}$ is divided into a coronal region (0 \textless\ $\theta$ \textless\ $\theta_{d}$) and a disk region ($\theta_d$ \textless\ $\theta$ \textless\ $\pi/2$) where, again, $\theta$ is the colatitude measured from the axisymmetry axis. The disk height at the boundary is set by satisfying hydrostatic equilibrium. Starting at \ttim\ = 0, we permit low-temperature, high-density matter to flow in through the disk boundary region ($\theta$ \textgreater\ $\theta_d$) with a fixed density  profile set by $\trho_{\rm d}$. The matter velocity along the disk boundary is set to be slightly sub-Keplerian so that the matter will flow into the simulation region. All other hydrodynamic variables have the ``free'' boundary conditions in the disk region, $\partial(\dots)/\partial r$ = 0.

In the coronal region ($\theta$ \textless\ $\theta_d$), the boundary conditions are also free for all hydrodynamic variables. However, we prohibit matter from flowing into the simulation region from this portion of the boundary. We solve the transport equation for the flux function $\Psi$ so that the magnetic flux flows out of the region together with matter.

The boundary conditions on the equatorial plane and on the rotation axis are symmetric and axisymmetric, respectively. On the axisymmetry axis, we  enforce the axisymmetry boundary conditions by requiring that the toroidal variables obey $B_\phi(-\theta) = -B_\phi(\theta)$, $v_\phi(-\theta) = -v_\phi(\theta)$, et cetera. This imposes the explicit requirement that the toroidal variables must cross 0 on the axis. Nonetheless, we must note that there is a region near the axisymmetry axis ($\theta = 0$ to $\theta_c$) where the azimuthal angular velocity $v_\phi/r$  may take on spurious values. This can in turn result in spurious values of $B_\phi$ near the axis even though $B_\phi(\theta=0)=0$. However, as we will discuss in \S \ref{subsec_properties}, the anomalous values do not affect the collimation of the jet. In our simulations $\theta_c$ spans about two angular grid cells or about $3.6^\circ$. 

\paragraph*{Internal boundary on the star}
The inner boundary $R_{\rm in} = R_*$ lies on the stellar surface. We assume that the poloidal component of the magnetic field $\bvec_p$ is frozen to the surface of the star: $B_r$ is held fixed while $B_\theta$ and $B_\phi$ obey the free boundary conditions $\partial B_\theta/\partial r = 0$ and $\partial B_\phi/\partial r = 0$. The plasma pressure and entropy also obey free boundary conditions on the stellar surface; at each timestep, the density on the surface is recalculated from these two free variables. The matter velocity components are calculated using free boundary conditions and then adjusted to be parallel to the magnetic field vectors on the stellar surface. We do not consider outflows due to stellar winds and therefore we only permit matter to flow inward onto the star.

\section{Simulation overview} \label{sec_overview}
\begin{figure*}[b]
\centering
\includegraphics[width=160mm]{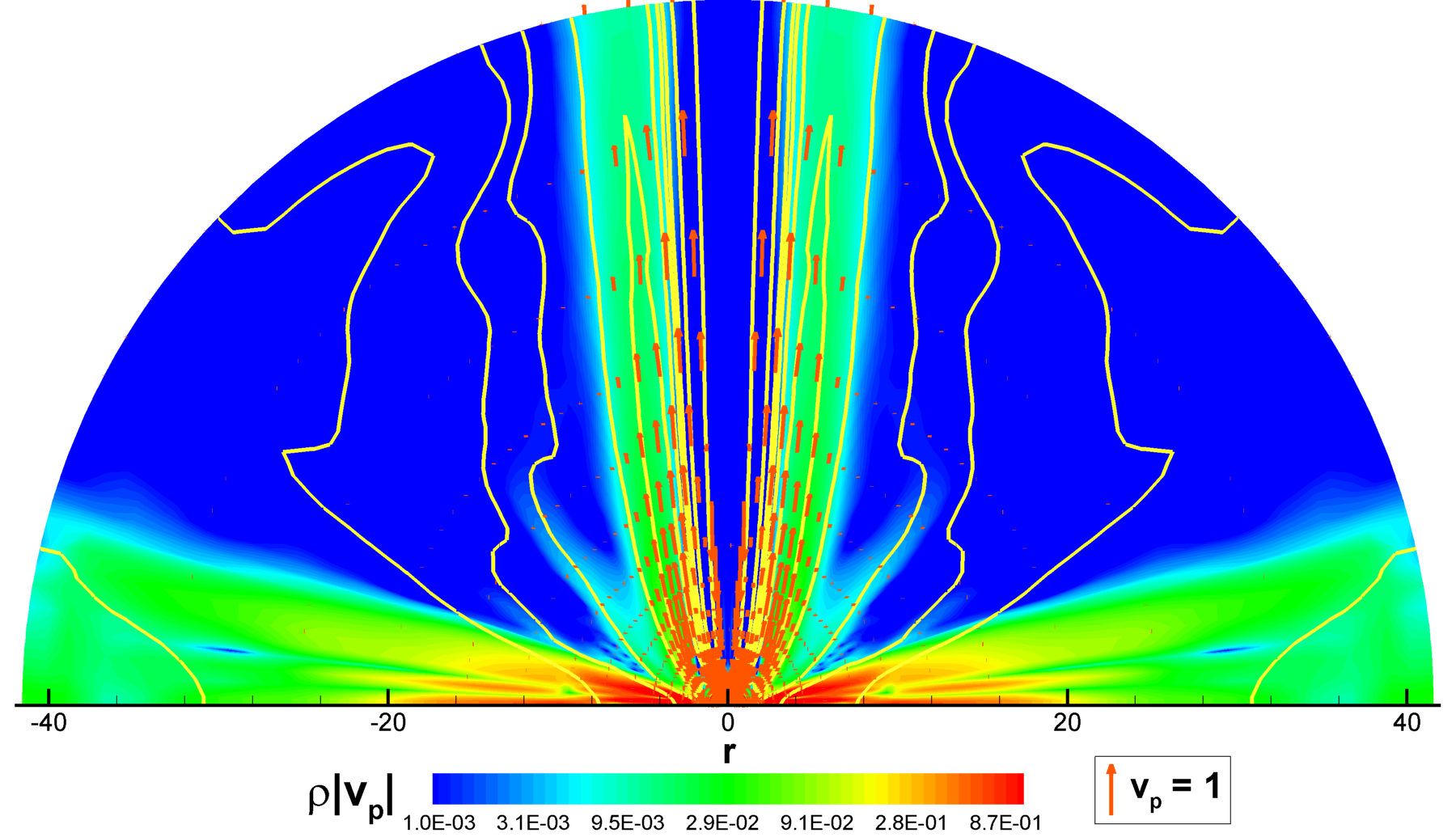}
\caption{The jet at time \ttim\ = 860. The background shows the poloidal matter flux density \rhov\ and the lines show contours of the magnetic flux function $\Psi$ which is a proxy for the poloidal magnetic field. The horizontal axis shows the distance away from the star in units of the reference radii $R_0$. The red vectors show the poloidal matter velocity ${\mathbf v_p}$. To obtain dimensional values, multiply these numbers by the numbers given in Tab. \ref{tbl_refval}. For a CTTS, $t_0$ = 0.366 days and $R_0$ = 2\rsun (e.g. column 1 in Tab. \ref{tbl_refval}): in real units then, the time \ttim=860 corresponds to 315 days and the simulation region is 0.39 AU in radius. \label{fig_fullregion}}
\end{figure*}

\begin{figure*}
\centering
\includegraphics[width=175mm]{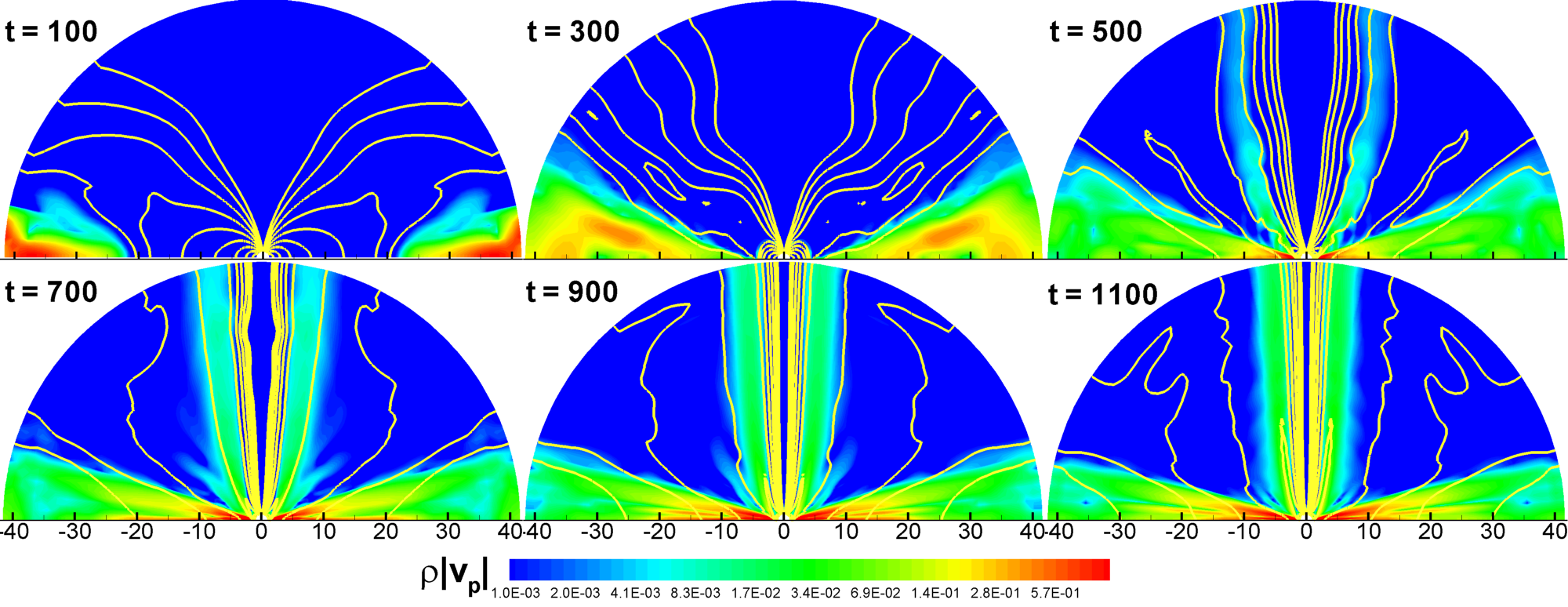}
\caption{Formation and collimation of the jet at different times \ttim. The \rhov\ background contour and field lines are plotted on the same scale as the contours in Fig. \ref{fig_fullregion}. \label{fig_timestep}}
\end{figure*}

In order to investigate formation of jets from the disk-magnetosphere boundary, we performed dozens of simulations with varying initial parameters and grid sizes. The region size is set by $N_\theta$ and $N_r$, the number of grid cells in the $\theta$ and r directions, respectively. The simulations have a total of 5 free parameters:  $\alpha_\nu$ and $\alpha_d$, the dimensionless viscosity and diffusivity coefficients; $\trho_{\rm d}$ and $\trho_{\rm c}$, the boundary densities of the disk and corona; and $\mu$, the stellar magnetic dipole moment. A sample of the runs which successfully produced strongly collimated jet-like outflows is shown in Tab. \ref{tbl_runs}. 

\begin{table}
\centering
\begin{tabular}{lllllll}
\hline \hline
$N_\theta$, $N_r$ & $\alpha_v$ & $\alpha_d$ & $\trho_{\rm d}$ & $\trho_{\rm c}$ & $R_{\rm out}$ & $\mu$ \\
\hline
30, 55	&	0.3	&	0.1	&	5	&	$10^{-3}$	&	16.5	&	3	\\ \hline
30, 66	&	0.3	&	0.1	&	0.1	&	$10^{-4}$	&	27	&	10	\\
30, 66	&	0.3	&	0.1	&	1	&	$10^{-3}$	&	27	&	10	\\
30, 66	&	0.3	&	0.1	&	10	&	$10^{-3}$	&	27	&	10	\\
30, 66	&	0.3	&	0.1	&	1	&	$10^{-3}$	&	27	&	10	\\
30, 66	&	0.6	&	0.3	&	10	&	$10^{-3}$	&	27	&	10	\\ \hline
30, 80	&	0.3	&	0.1	&	10	&	$10^{-3}$	&	58	&	10	\\
30, 80	&	0.6	&	0.3	&	10	&	$10^{-3}$	&	58	&	10	\\ \hline
50, 100	&	0.3	&	0.1	&	5	&	$10^{-3}$	&	22	&	10	\\
50, 100	&	0.3	&	0.1	&	10	&	$10^{-3}$	&	22	&	10	\\ \hline
50, 120	&	0.3	&	0.03	&	5	&	$10^{-3}$	&	42	&	10	\\
50, 120	&	0.3	&	0.1	&	3	&	$10^{-3}$	&	42	&	10	\\
50, 120	&	0.3	&	0.1	&	5	&	$10^{-3}$	&	42	&	3	\\ 
{\bf 50, 120}	&	{\bf 0.3}	&	{\bf 0.1}	&	{\bf 5}	&	{\bf 10$^{-3}$}	&	{\bf 42}	&	{\bf 10}	\\
50, 120	&	0.3	&	0.1	&	5	&	$10^{-3}$	&	42	&	15	\\
50, 120	&	0.3	&	0.1	&	5	&	$10^{-3}$	&	42	&	25	\\
50, 120	&	0.3	&	0.1	&	10	&	$10^{-3}$	&	42	&	10	\\\hline \hline
\end{tabular}
\caption{Parameters for simulation runs which produced well collimated jets. $N_\theta$ and $N_r$ are the number of grids in the $\theta$ and radial directions, respectively; $\alpha_v$ and $\alpha_d$ are the dimensionless viscosity and diffusivity coefficients. In dimensionless units, $\trho_{\rm d}$ is the fixed density on the disk boundary; $\trho_{\rm c}$ is the fixed coronal density at the external boundary; $R_{\rm out}$ is the radius of the outer simulation boundary (this is set by $N_\theta$ and $N_r$, so it is not a free parameter), and $\mu$ is the stellar magnetic dipole moment. The bold line is the reference case which we use for analysis in this paper. \label{tbl_runs}}
\end{table}

We use the Keplerian rotation period at the surface of the star $t_0$ as a reference timescale. We observe three general classes of outflows: one-time episodic outflows, stable weakly collimated conical-type outflows, and stable collimated jets. The one-time episodic outflows occur when the accreting matter strongly compresses the stellar field lines, driving transient outflows from the disk-magnetosphere boundary. In this case, the matter and magnetic pressures quickly re-equilibrate, halting the driving mechanism and suppressing the outflow in just a few $t_0$. The conical winds are similar to those previously studied in R09 and are driven by the compressed magnetic field at the disk-magnetosphere boundary. In this work, we are interested in particular in the last class of observed outflow: the stable collimated jet.


\subsection{Reference run}
\begin{figure*}
\centering
\includegraphics[width=160mm]{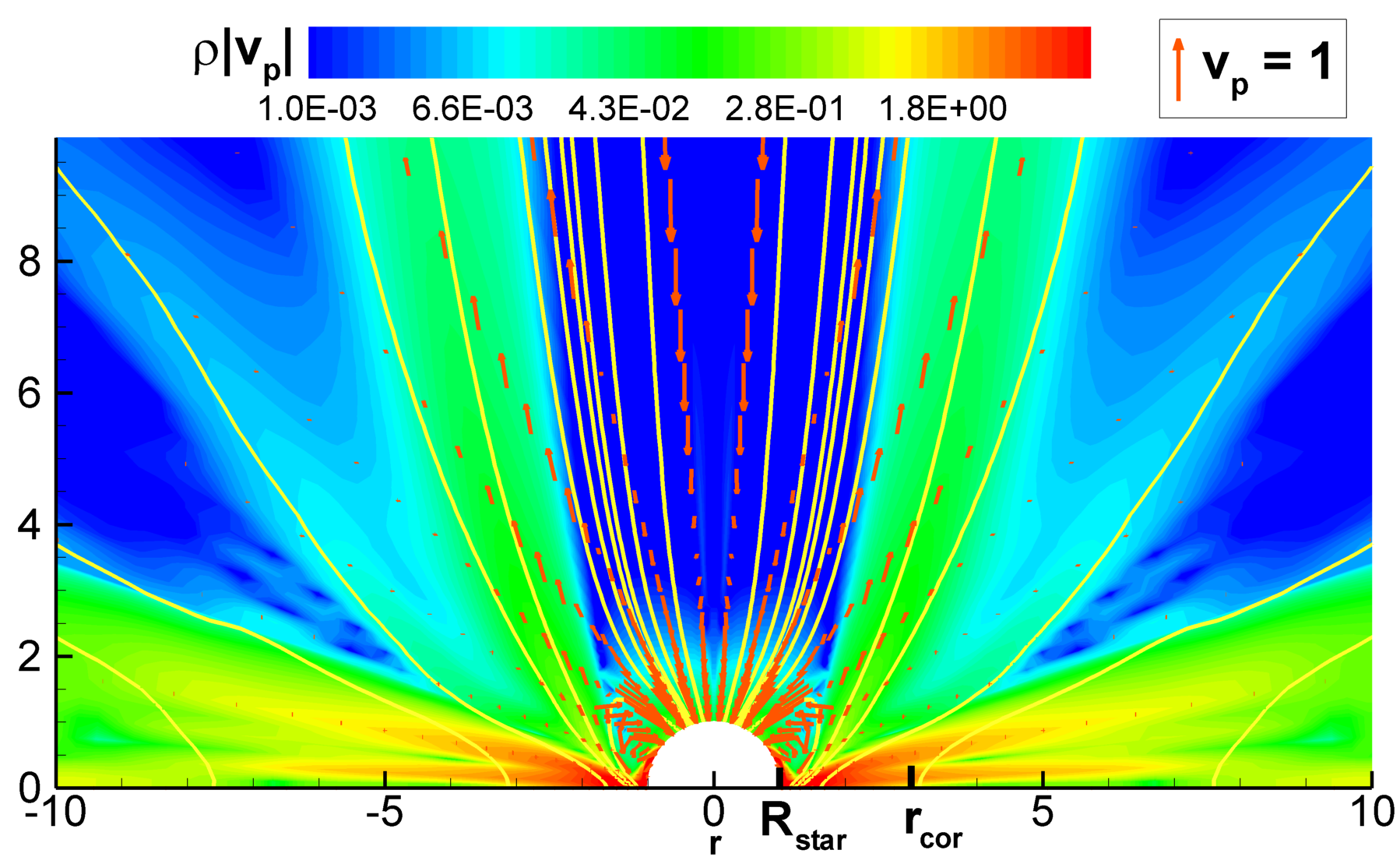}
\caption{A zoomed-in plot of the inner disk region at the time \ttim=860. The plotted contours show the same scales as in Fig. \ref{fig_fullregion}. \label{fig_zoomregion}}
\end{figure*}
Fig. \ref{fig_fullregion} shows one of the simulations which produced a strongly collimated jet: throughout the remainder of the paper, we will use this run as a reference case to perform additional analysis. The background contours represent the poloidal matter flux \rhov\ and the lines represent contours of the magnetic flux function $\Psi$, which obeys the relation $\bvec \cdot \nabla\psi = 0$. In the poloidal plane, therefore, the $\psi$ contours represent the magnetic field lines. The reference simulation uses a spherical grid with $N_\theta$ = 50 cells in the $\theta$ direction and $N_r$ = 120 cells in the radial direction with the parameters: $\alpha_v = 0.3$, $\alpha_d = 0.1$, disk density $\trho_{\rm d} = 5$, and coronal density $\trho_{\rm c} = 0.001$ (a summary of the parameters is shown as the bolded line in Tab. \ref{tbl_runs}). For analysis of the launching and collimation mechanisms, we study the simulation at \ttim\ = 860 when the jet is well established.

Fig. \ref{fig_timestep} shows snapshots of the reference simulation at six moments in time \footnote{An animation of the full reference simulation is available at http://astro.cornell.edu/$\sim$pslii/research.htm. We encourage the reader to refer to the online animations for clarification of the discussion in this section.}. At \ttim\ = 100, the matter flowing in from the boundary begins compressing and inflating the dipole field around the star. 
By \ttim\ = 300, the disk is only a few stellar radii away from the star; the close proximity of the disk permits matter to accrete directly onto the star by flowing along the closed dipole field lines. At the same time, the inflated field lines in the corona undergo forced reconnection, ejecting plasmoids into the stellar corona. 

The strong magnetic gradient at the disk-magnetosphere boundary launches matter along the inflated dipole field lines. The outflow first emerges from the open field lines at the disk-magnetosphere boundary starting around \ttim\ = 350 and becomes stable and well established by \ttim\ = 500. As the jet stabilizes, the magnetic reconnection within the jet halts and the matter flows smoothly. Fig. \ref{fig_zoomregion} shows a zoomed-in view of the outflow at \ttim\ = 860, illustrating the proximity of the launching region to the star. As the outflow becomes stronger, we observe a gradual magnetic collimation of the outflowing matter into a jet (see frames \ttim\ = 700 to \ttim\ = 1100 of Fig. \ref{fig_timestep}).

\begin{figure}
\centering
\includegraphics[width=50mm]{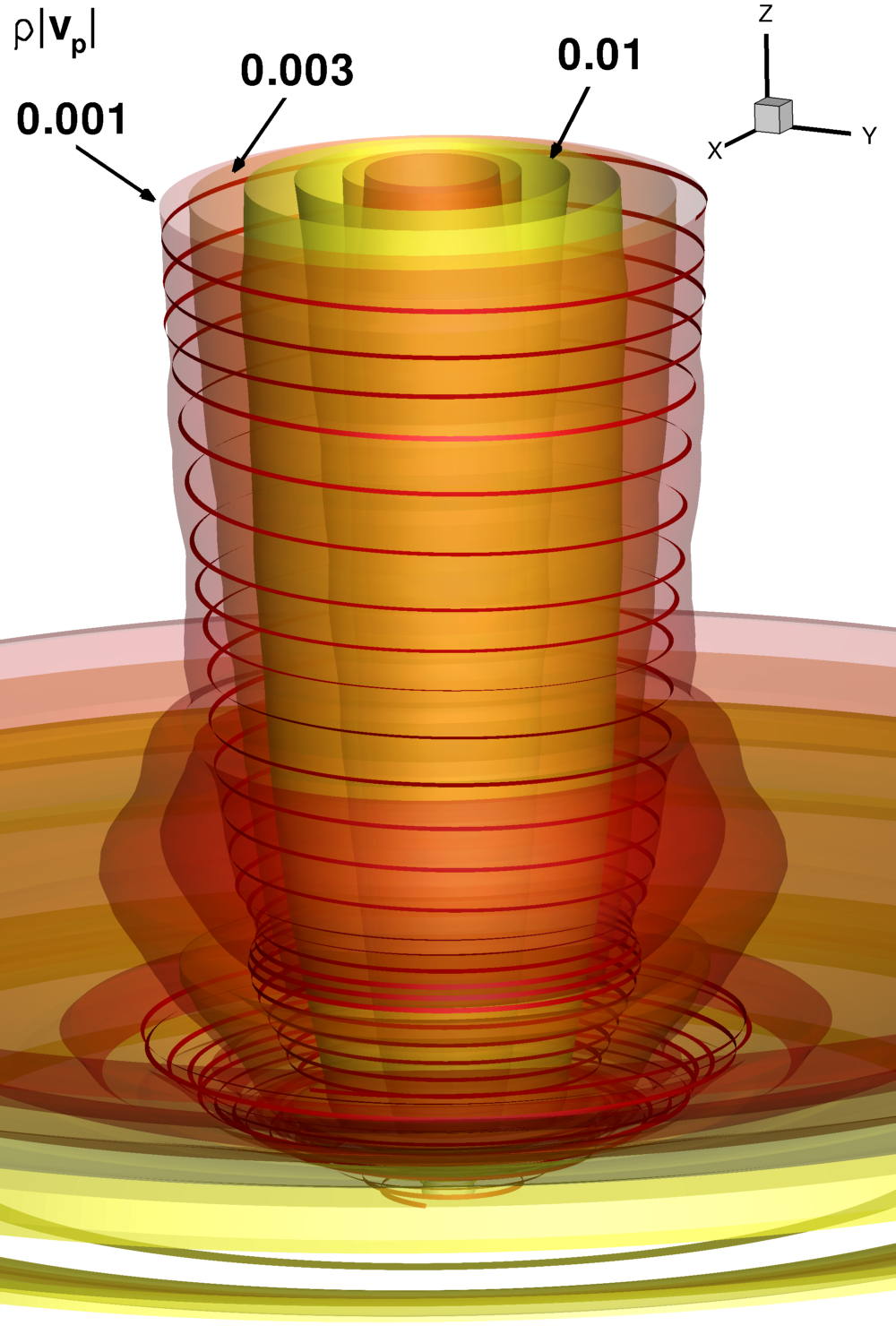}
\caption{3D view of the poloidal matter flux density \rhov\ contours of the jet at \ttim\ = 860. Three matter flux density surfaces are plotted: \rhov\ = 0.001, 0.003, and 0.01. The red streamtraces show the strongly wound magnetic field lines in the corona of the star which collimate the outflowing matter. \label{fig_3dregion}}
\end{figure}

In Fig. \ref{fig_3dregion}, we show a three-dimensional rendering of the jet at \ttim\ = 860. Three poloidal matter flux surfaces are shown along with tightly wound helical magnetic field lines---represented by the red ribbons---which play a key role in collimating the outflow (we discuss this later in \S\ref{subsec_forces}). The jet has the shape of an inverted tapered cone with a cylindrical radius of $1-4\ R_*$ at the base of the jet and $3-8\ R_*$ at the top. For this reference case, we measure a half-opening angle of $\Theta \approx 4^\circ$ at the top of the simulation region.


\subsection{Properties of the jet} \label{subsec_properties}
\begin{figure}
\centering
\includegraphics[width=85mm]{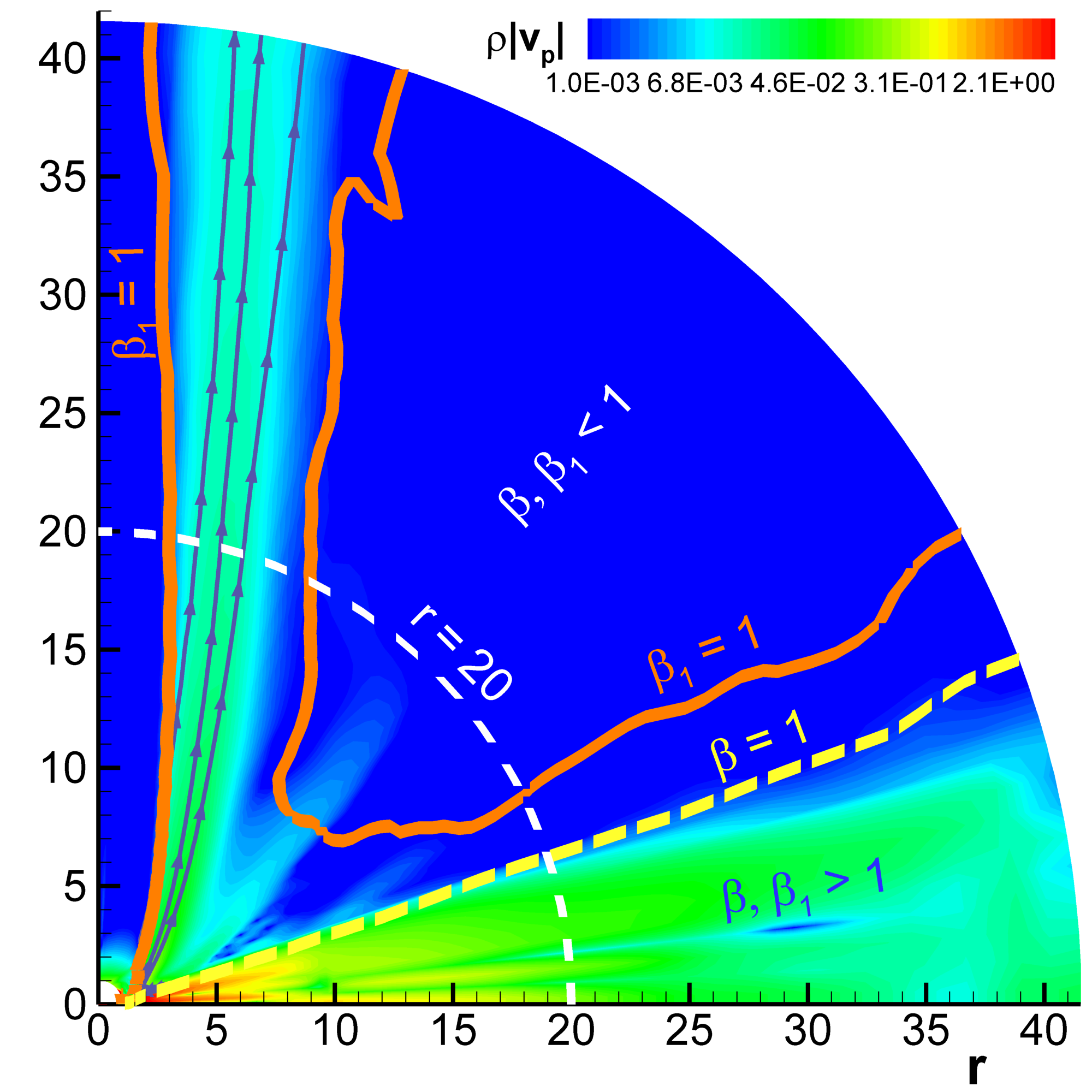}
\caption{$\rhov$ contours overplotted with the $\beta = 1$ and $\beta_1 = 1$ lines (see Eq. \ref{eqn_beta}). The region interior to the $\beta_1$ line is matter dominated. The dashed white line shows the radius where the $\theta$ cross-sections shown in Fig. \ref{fig_theta} are taken. 
Also plotted are a few of the velocity streamlines in the jet. \label{fig_regionbeta}}
\end{figure}
In order to describe the relative magnitudes of the matter and magnetic pressures around the star, we use the standard plasma parameter $\beta$ and the kinetic plasma parameter $\beta_1$:
\begin{align} 
	\beta = \frac{P}{B^2 / 8 \pi}  \label{eqn_beta}, & & \beta_1  = \frac{P + \rho {\bf v}^2}{B^2 / 8 \pi}.
\end{align}
The kinetic plasma parameter $\beta_1$ takes into account the ram pressure of the gas, $\rho{\bf v}^2$ in addition to the thermal gas pressure, $P$. Regions with $\beta, \beta_1 \gg$ 1 are matter pressure dominated while regions with $\beta, \beta_1 \ll$ 1 are magnetic pressure dominated. Fig. \ref{fig_regionbeta} shows the $\beta$, $\beta_1 = 1$  lines for the \ttim\ = 860 reference case. The conventional plasma parameter $\beta$ is much less than unity everywhere except within the disk; hence if we use the plasma $\beta$ criterion, we find that the jet is completely magnetically dominated. However, if the ram pressure is taken into account and $\beta_1$ is used instead, then we find that $\beta_1 \apprge 1$ inside the jet: in other words, the magnetic pressure is only a few times smaller than (or comparable to) the effective matter pressure, showing that the magnetic field is important in driving and collimating the jet even at large distances from the star.

\begin{figure*}
\centering
\includegraphics[width=180mm]{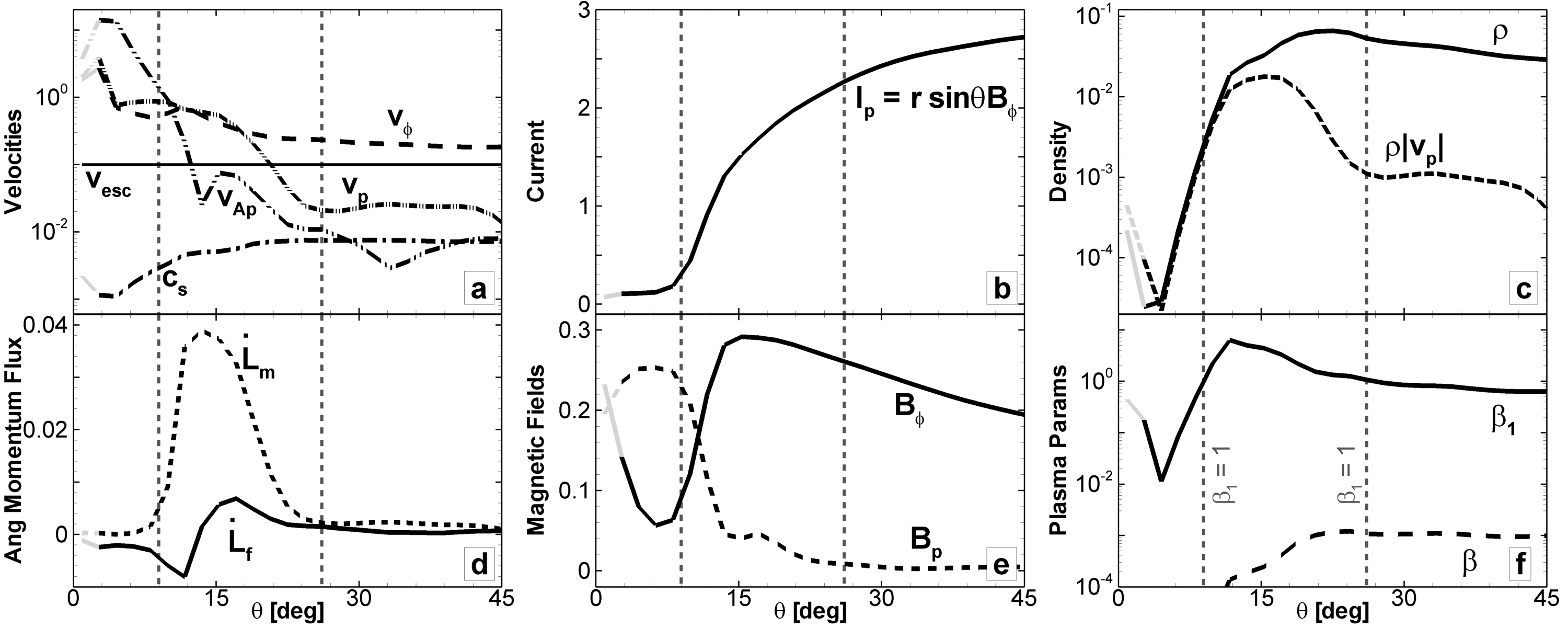}
\caption{Cross-sections of the jet in the $\theta$ direction at \trad\ = 20 and \ttim\ = 860 (see Fig. \ref{fig_regionbeta}). The dashed vertical lines indicate the $\beta_1$=1 surfaces which approximately represent the boundaries of the jet. We indicate the region where numerical artifacts may play a role by greying out the two grids closest to the axis. {\bf Panel (a)} shows the velocity cross-section in the jet. $v_\phi$ and $v_p$ are the velocities in the toroidal and poloidal directions, respectively; $v_{\rm esc}$ is the escape velocity at \trad\ = 20; $c_s$ is the sound speed; and $v_{\rm Ap}$ is the poloidal Alfv$\acute{\rm e}$n speed. {\bf Panel (b)} shows the poloidal current $I_p$ in the jet. {\bf Panel (c)} shows the density $\rho$ and matter flux $\rhov$ profiles in the jet. {\bf Panel (d)} shows $\ldot_m$ and $\ldot_f$, the angular momentum fluxes from matter and magnetic fields, respectively. {\bf Panel (e)} plots the magnetic field components in the poloidal $B_p$ and toroidal $B_\phi$ directions. And lastly, {\bf Panel (f)} shows the plasma $\beta$ parameters in the jet (described in Eq. \ref{eqn_beta}). \label{fig_theta}}
\end{figure*}
Fig. \ref{fig_theta} shows cross-sections of the jet in the $\theta$ direction for angles $\theta$ = $0^\circ$--$45^\circ$ at the \trad\ = 20 surface (see Fig. \ref{fig_regionbeta}). The dashed vertical lines represent the $\beta_1$ = 1 surfaces which {\it very approximately} delimit the boundaries of the jet. We indicate the region where numerical artifacts may play a role by greying out the two grids closest to the axis. Fig. \ref{fig_theta}f shows the $\beta$ and $\beta_1$ profiles of the jet, with $\beta \approx 10^{-4}-10^{-3}$ and $\beta_1 \approx 1-6$ inside, indicating that the jet is ``weakly'' matter dominated. Fig. \ref{fig_theta}a shows the velocity cross-section of the jet: within the jet, the matter is supersonic and the poloidal matter velocity $\tvel_p$ is larger than the poloidal Alfv$\acute{\rm e}$n speed $\tvel_{\rm Ap} = |{\bf B}_p|/\sqrt{4\pi\rho}$, except at the left edge of the jet where they are comparable. The poloidal velocities in the jet range from $\tvel_p \approx 1$ on the inner edge of the jet to $\tvel_p \approx 0.1$ on the outer edge. 

Panel \ref{fig_theta}a also shows that the toroidal matter velocity $v_\phi$ is very large even outside the jet, contributing to the winding of the magnetic field lines in the corona. Fig. \ref{fig_theta}b shows the profile of the normalized poloidal current $I_p = r \sin\theta B_\phi$ flowing out of the simulation region through the jet (see Appendix \ref{subsec_current} for analysis of the current flow). Figs. \ref{fig_theta}c and \ref{fig_theta}d show the matter and angular momentum flux profiles in the jet: we see that the angular momentum transport is dominated by the outflowing matter\footnote{In contrast, the opposite is true at small radii ($\trad\ \lesssim 5$) and angular momentum is primarily transported away by the magnetic field.} (this is further discussed in \S\ref{subsec_fluxes}). Panel \ref{fig_theta}e shows the magnetic field distribution; note that the toroidal field is much larger than the poloidal field everywhere except on the inside edge of the jet, where we observe spurious values of $B_\phi$. Note, however, that the magnetic collimation force is proportional to the transverse gradient of $I_p^2 = (r \sin\theta B_\phi)^2$ which is well behaved near the symmetry axis, as shown in Fig. \ref{fig_theta}b, and not on $B_\phi^2$ itself \citep{lovelace1989}. Therefore, the spurious axis values of $B_\phi$ do not affect the collimation of the jet. Rather, the dominant collimating force arises far from the axis between $\theta \sim 5-15^\circ$ where both $B_\phi$ and $I_p$ are well behaved. In fact, the gradients of $(r \sin\theta B_\phi)^2$ and ${\bf B}_{\rm p}^2$ both contribute to the magnetic force and in \S \ref{subsec_forces}, we discuss in detail how these magnetic fields launch and collimate the outflow.

\section{Fluxes, forces, and velocities in the jet} \label{sec_analysis}
\begin{figure*}
\centering
\includegraphics[width=150mm]{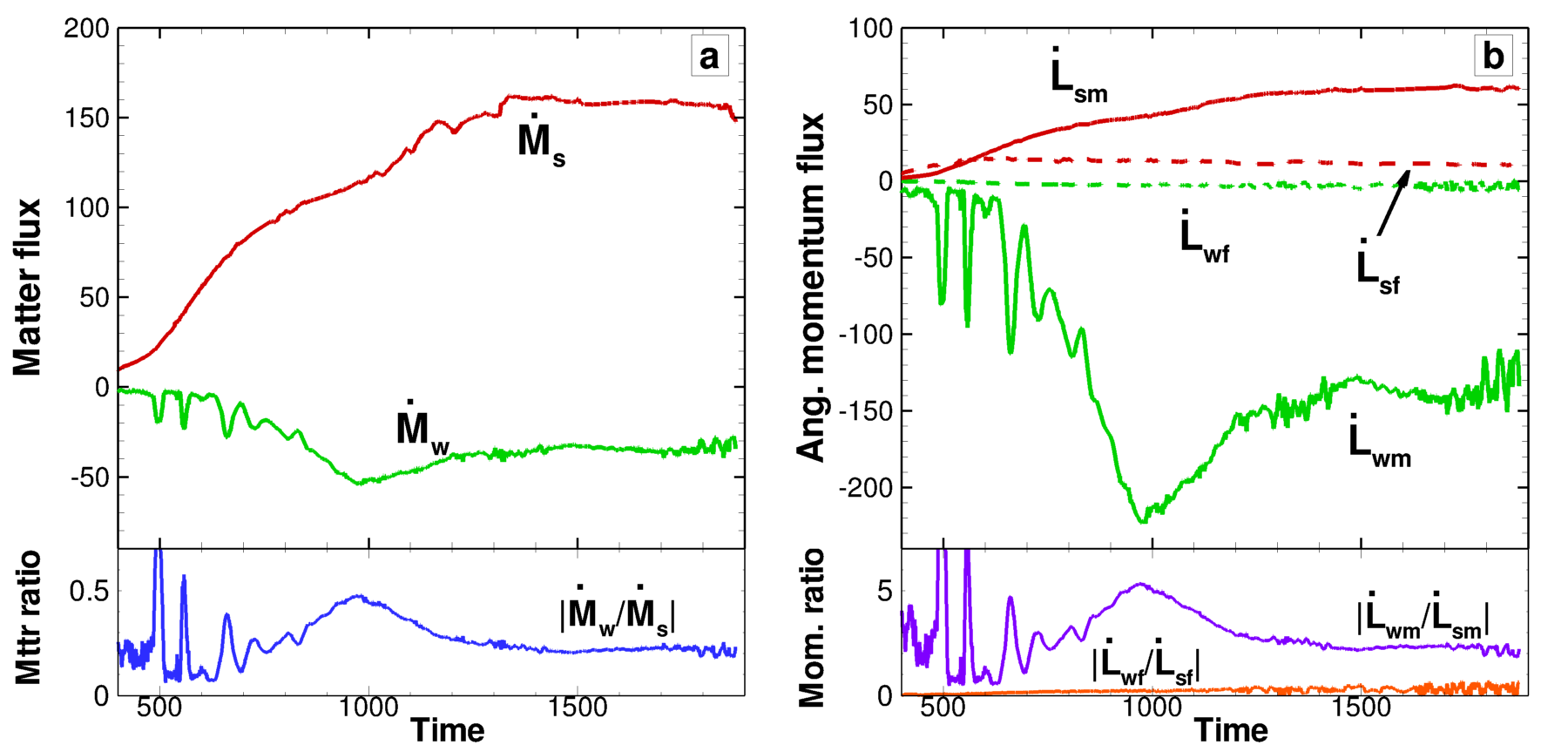}
\caption{Matter and angular momentum fluxes onto the star and into the outflow. {\bf Panel (a)}: Matter flux from the disk onto the star $\mdot_s$ and into the jet ${\mdot}_w$ as a function of time. The bottom plot shows the absolute value of the accretion-to-ejection ratio $|\mdot_w/\mdot_s|$ as a function of time. {\bf Panel (b)}: Angular momentum flux rates added onto the star by matter $\ldot_{sm}$ and by the magnetic fields $\ldot_{sf}$. Also plotted is the angular momentum loss rate due to matter carried away by the jet $\ldot_{wm}$ and by the magnetic fields $\ldot_{wf}$. The bottom plot shows the ratios of angular momentum fluxes leaving in the jet to the angular momentum fluxes onto the star.  \label{fig_flux}}
\end{figure*}
\subsection{Matter and angular momentum flux} \label{subsec_fluxes}
\subsubsection{Matter flux}
Here we analyze the matter and angular momentum fluxes integrated over the star's surface (\trad\ = 1) and the \trad\ = 20 surface for the reference simulation. The mass accretion rate through a given surface is
\begin{equation} \label{eqn_mdot}
\mdot = \int d {\bf S}\cdot \rho {\mathbf v_p} = \int dS\rho v_r~,
\end{equation}
where $d {\bf S}$ is an outward facing surface area element and $v_r$ is the radial component of the velocity. On the stellar surface, we measure the accretion rate $\mdot_{\rm s}$\ by only considering the inward ($v_r < 0$) matter flux; conversely, for the outer \trad\ = 20 surface we measure the outflow rate in the jet $\mdot_{\rm w}$\ by only considering the outward $v_r > 0$ matter flux. Fig. \ref{fig_flux}a shows the inward $\mdot_{\rm s}$\ across the stellar surface and outward $\mdot_{\rm w}$\ across the \trad\ = 20 surface as a function of time. The total matter flux onto the star $\mdot_{\rm s}$ slowly increases from the time the disk first reaches the star (\ttim\ = 300) until it stabilizes around \ttim\ = 1300. Similarly, the total matter flux through the jet stabilizes slightly earlier at \ttim\ = 1200. 
\subsubsection{Angular Momentum Fluxes}
The angular momentum flux density consists of three components: angular momentum carried by the matter ${\bf F}_{\rm Lm}$, by the magnetic field ${\bf F}_{\rm Lf}$, and by the viscosity ${\bf F}_{\rm Lv}$. The total angular momentum flux density is therefore
$${\bf F}_{\rm L}={\bf F}_{\rm Lm}+{\bf F}_{\rm Lf}+{\bf F}_{\rm Lv},$$
where
\begin{align}\label{eqn_flux}
{\bf F}_{\rm Lm} & =  r \sin \theta \rho v_\phi  {\bf v}_p~,\nonumber \\
{\bf F}_{\rm Lf} & = -r \sin \theta \frac{B_\phi {\bf B}_p}{4 \pi}~, \\
{\bf F}_{\rm Lv} & = - \nu_t \rho (r \sin \theta)^2 {\bf \nabla}\Omega~, \nonumber
\end{align}
with the last term non-zero only inside the disk. As with the matter flux, we integrate the angular momentum fluxes at the star (\trad\ = 1) and at the \trad\ = 20 surface,
\begin{equation}\label{eqn_ldot}
\ldot = \int d {\bf S}\cdot {\bf F}_{\rm L} = \int d {\bf S}\cdot ({\bf F}_{\rm Lm} + {\bf F}_{\rm Lf} + {\bf F}_{\rm Lv}).
\end{equation}
Fig. \ref{fig_flux}b shows the integrated angular momentum flux as a function of time. $\ldot_{sm}$ and $\ldot_{sf}$ measure the angular momentum being {\it added} to the star by matter and magnetic fields, respectively. The accreting matter adds angular momentum directly ($\ldot_{sm}$) and spins up the star. The magnetic fields also allow the matter to spin up the star indirectly: the magnetospheric radius $\trad_{m} \approx 1.1$ (calculated by equating the magnetic and gas pressure in the disk) is smaller than the corotation radius $r_{cor} = 3$ and hence the inner disk drags the stellar dipole field, causing the star to spin up at a nearly constant rate ($\ldot_{sf}$). 

At the \trad\ = 20 surface, only the outward angular momentum carried by the jet is calculated. The outward angular momentum transport due to magnetic fields, $\ldot_{\rm wf}$, is negligible compared to the the angular momentum being carried away by the matter, $\ldot_{\rm wm}$. Like the matter flux, the angular momentum flux in matter becomes steady around \ttim\ = 1200. Roughly a third of the disk's angular momentum exits the system through the jet, while only a small fraction of the incoming angular momentum is added to the star. The majority of the disk's angular momentum is transported by viscosity back and out through the disk itself.

\subsection{Velocities in the jet} \label{subsec_veloc}
\begin{figure}
\centering
\includegraphics[width=85mm]{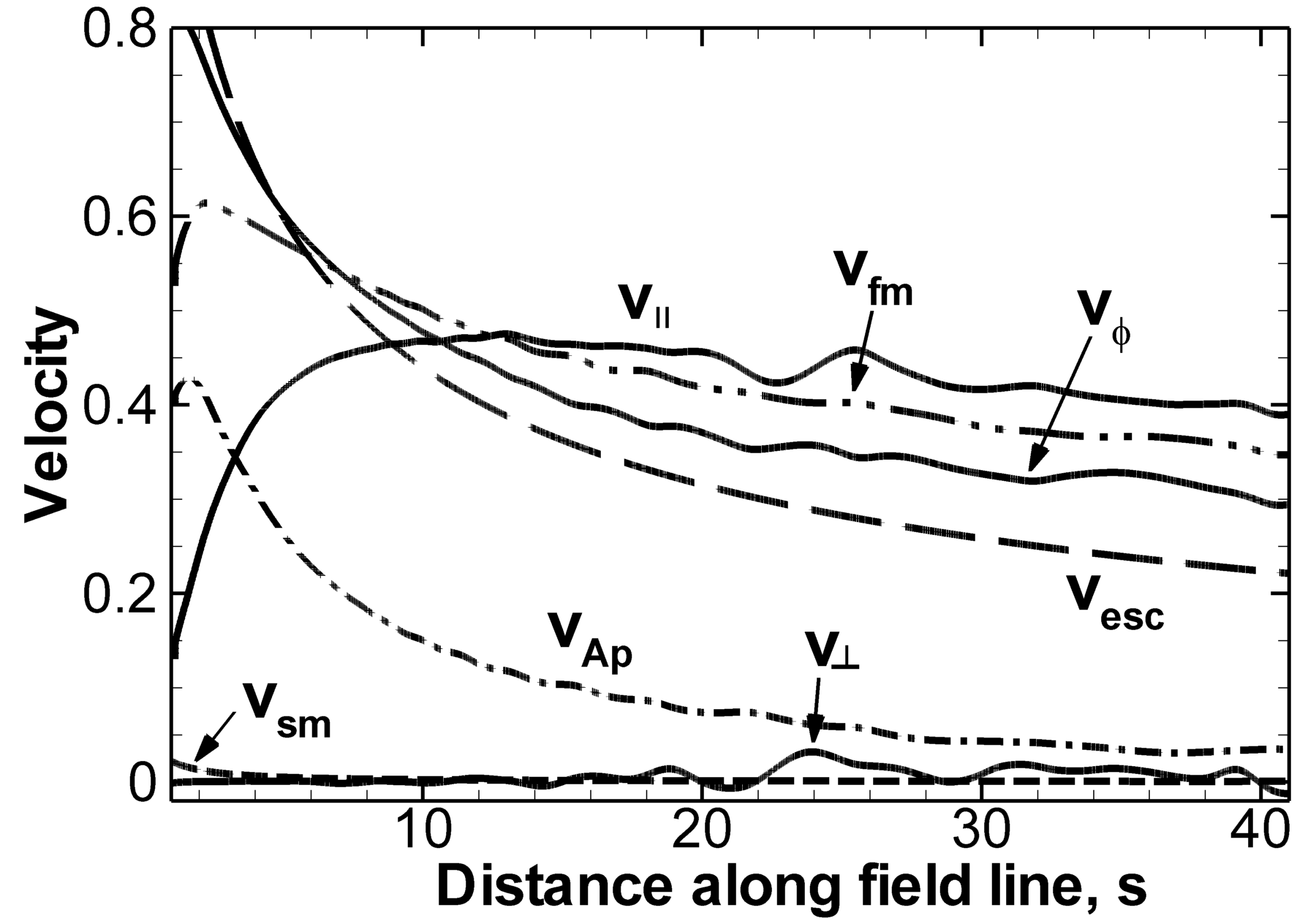}
\caption{Velocities as a function of distance along a representative field line. The field line is anchored in the disk at \trad=1.24 and extends through the jet to the outer boundary of the simulation region. A total of eight velocities are plotted. Starting from the left:  $v_{\rm sm}$ - slow magnetosonic velocity, $v_{Ap}$ - poloidal Alfv$\acute{\rm e}$n velocity, $v_{\parallel}$ - velocity of matter tangential to the field line, $v_{\perp}$ - velocity of matter perpendicular to the field line, $v_{\rm fm}$ - fast magnetosonic velocity, $v_\phi$ - toroidal matter velocity along the field line, and lastly, $v_{\rm esc}$ - the escape velocity. \label{fig_velocity}}
\end{figure}
Fig. \ref{fig_regionbeta} shows a few of the velocity streamlines flowing from the base of the jet to the edge of the simulation region. The figure shows that the launching region of the jet is localized to the very inner disk. To probe the structure of the jet in detail, we analyze the velocities within the jet at \ttim\ = 860. Fig. \ref{fig_velocity} shows velocities along a ``representative'' field line running through the jet in the \ttim\ = 860 reference case. The field line originates from the disk at \trad\ = 1.24 
and extends upward through the jet to the outer boundary of the simulation region (for a visual representation, see the denoted field line in Fig. \ref{fig_streamline}). 
We define the new coordinate \ts\ which traces the linear distance along the poloidal field line with \ts\ = 0 at the base\footnote{Since the representative field line is nearly vertical, $\ts\approx\trad\cos\theta$}. Fig. \ref{fig_velocity} shows that very close to the star, $v_{\parallel}$---the matter velocity tangential to the field line---is small; however, $v_{\parallel}$ quickly rises above the slow magnetosonic $v_{\rm sm}$, poloidal Alfv$\acute{\rm e}$n $v_{\rm Ap}$, and fast magnetosonic $v_{\rm fm}$ speeds as the matter moves away from the star, indicating that the flow is matter dominated past \ts\ $\approx$ 13; most of the acceleration occurs before the matter crosses the fast magnetosonic surface. Fitting a power law to the region of strong accleration gives $v_{\parallel} \propto s^{0.71}$; past $\ts\ \sim 5$, the acceleration mechanism weakens and the slope flattens such that $v_{\parallel} \propto s^{0.16}$. Near the disk, the toroidal component of the velocity, $v_\phi$, is close to the Keplerian velocity meaning that the matter in the jet initially corotates with the disk. However, outside the disk, $v_\phi$ falls off as $s^{-0.33}$ and $v_{\parallel}$ exceeds $v_{\phi}$ at \ts\ $\approx 12$ due to the continued acceleration of the matter.

Far away from the star, the tangential velocity is nearly double the escape velocity $v_{\rm  esc}$ and the matter in the jet easily escapes from the system.  In contrast, the perpendicular velocity $v_{\perp}$ in the jet is quite small, indicating that matter diffuses across the magnetic field very slowly. The small oscillations perpendicular to the field line may be an Alfv$\acute{\rm e}$n wave, but this remains to be investigated.

\subsection{Launching and collimation mechanisms} \label{subsec_forces}
\subsubsection{Forces in the jet} \label{subsubsec_forceeqns}
\begin{figure*}
\centering
\includegraphics[width=170mm]{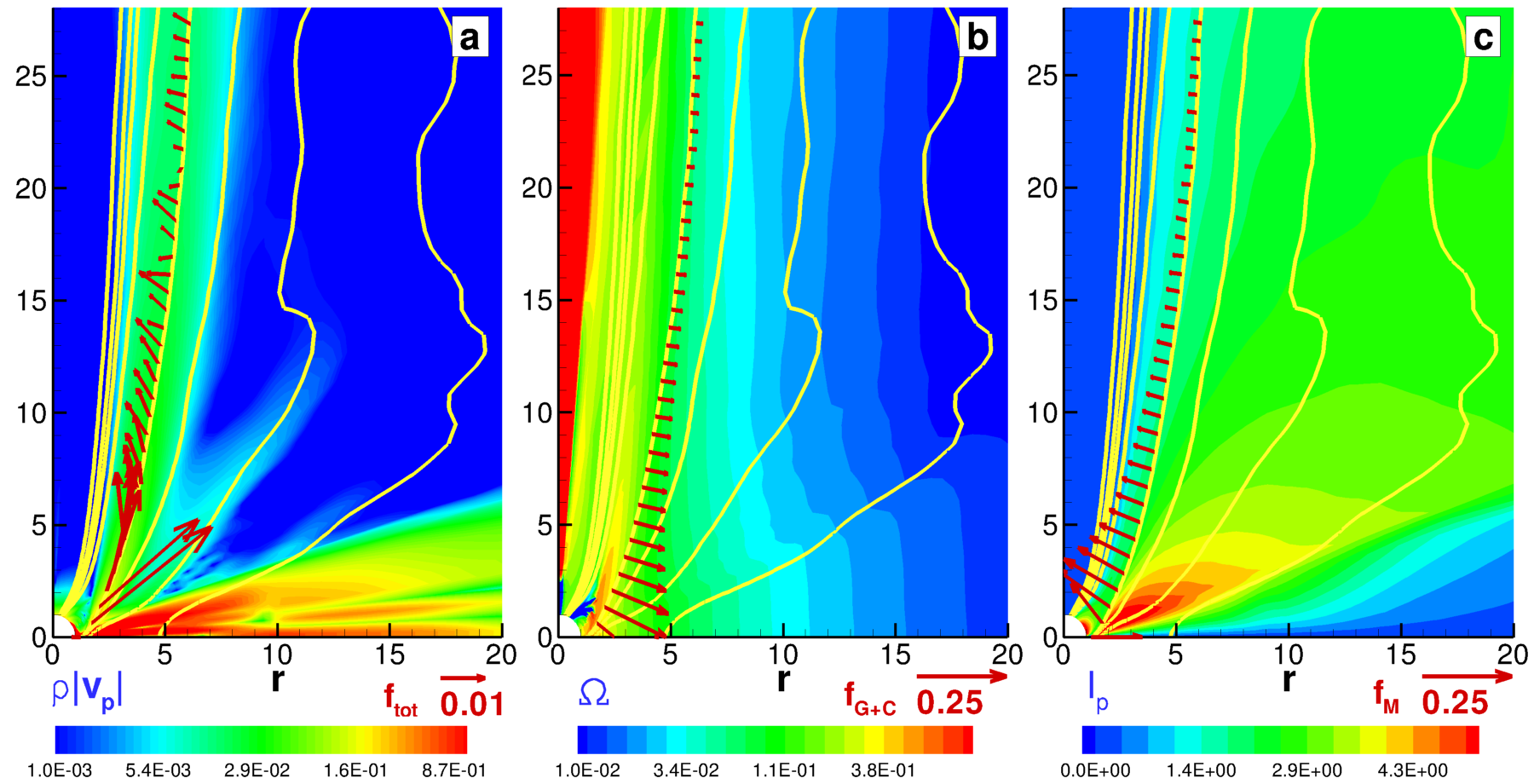}
\caption{Forces along a field line in the jet.
{\bf Panel (a)} shows the poloidal matter flux density \rhov\ as a background overplotted with poloidal magnetic field lines. The vectors show the {\it total} force ${\mathbf f}_{\rm tot}$ along a representative field line originating from the disk at \trad\ = 1.24. 
{\bf Panel (b)} plots the angular velocity $\Omega$ as the background. The vectors show the sum of the {\it gravitational + centrifugal} forces ${\mathbf f}_{\rm G+C}$ along the representative field line.
{\bf Panel (c)} shows the poloidal current $I_p$ as the background. The vectors show the total {\it magnetic} force ${\mathbf f}_{\rm M}$ along the representative field line. \label{fig_streamline}} 
\end{figure*}
The force per unit mass tangent to a poloidal magnetic field line can be calculated by taking the dot product of the Euler equation with the $\hat{\mathbf b}$ unit vector which is parallel to the poloidal magnetic field line ${\mathbf B}_p$. Following the derivation in \citet{ustyugova1999}, the total force tangent to ${\bf B}_p$ is
\begin{align} 
f_{\rm tot, \parallel} & = \hat{\mathbf b} \cdot ({\bf f}_{\rm P} + {\bf f}_{\rm G} + {\bf f}_{\rm C} + {\bf f}_{\rm M}) \label{eqn_forcet} \\
 = & - \frac{1}{\rho}\frac{\partial P}{\partial s} - \frac{\partial \Phi}{\partial s} + \frac{v_\phi^2}{r\sin\theta} \sin\Theta + \frac{1}{4 \pi \rho} \hat{\mathbf b} \cdot [(\nabla \times \bvec) \times \bvec]. \nonumber 
\end{align}
Here, ${\bf f}_{\rm P}, {\bf f}_{\rm G}, {\bf f}_{\rm C}, {\bf f}_{\rm M}$ are the pressure, gravitational, centrifugal and magnetic forces, respectively; $\Theta$ is the angle between the poloidal magnetic field line and the axisymmetry axis; $\Phi$ is the gravitational potential; $s$ is the previously defined coordinate which traces the distance along the poloidal field line.

The pressure gradient force, ${f}_{\rm P, \parallel} = -(1/\rho)(\partial P/\partial s)$, dominates within the disk. The matter in the disk is approximately in Keplerian rotation such that the sum of the gravitational and centrifugal forces roughly cancel (${\bf f}_{\rm G+C} \approx 0$). Near the slowly rotating star, however, the matter is strongly coupled to the stellar magnetic field and the disk orbits at sub-Keplerian speeds, giving ${\bf f}_{\rm G+C} \lesssim 0$.
The tangential magnetic force (the last term of Eq. \ref{eqn_forcet}) can be expanded as 
\begin{align}
f_{\rm M, \parallel} & = \frac{1}{4 \pi \rho}\hat{\mathbf b} \cdot [(\nabla \times \bvec) \times \bvec] \nonumber \\
& = -\frac{1}{8 \pi \rho (r\sin\theta)^2}\frac{\partial (r\sin\theta B_\phi)^2}{\partial s}
\end{align}
\citep{lovelace1991}.
Note that $r\sin\theta B_\phi$ is the normalized poloidal current flowing through a surface of radius $r$ from colatitude zero to $\theta$. 

The force per unit mass {\it perpendicular} to a poloidal field line is given by
\begin{align}
f_{\rm tot, \perp} & = -(v_p^2 - v_{\rm Ap}^2) \frac{\partial \Theta}{\partial s} \nonumber \\
 & ~ - \frac{1}{\rho}\frac{\partial}{\partial n}\left(P + \frac{\bvec_p^2}{8\pi}\right) -\frac{1}{8 \pi \rho (r\sin\theta)^2}\frac{\partial (r\sin\theta B_\phi)^2}{\partial n} \nonumber \\
&~  + \frac{v_\phi^2}{r\sin\theta}\cos\Theta - \frac{\partial \Phi}{\partial n}~, \label{eqn_forcen}
\end{align}
where $n$ is a coordinate normal to the poloidal field line \citep{ustyugova1999}. In \S \ref{subsec_veloc} we showed that far from the disk, the poloidal velocity in the jet $v_{\rm p}$ dominates over the poloidal Alf$\acute{\rm e}$n speed $v_{\rm Ap}$ (see Fig. \ref{fig_velocity}), permitting us to ignore the poloidal Alfv$\acute{\rm e}$n term. Additionally, far outside the disk, the pressure gradient term is negligible and gravity is weak enough to be ignored. With these simplifications, Eq. \ref{eqn_forcen} reduces to
\begin{align}
f_{\rm tot, \perp} & = - v_{\rm p}^2 \frac{\partial \Theta}{\partial s}
 - \frac{1}{8\pi \rho}\frac{\partial{\bf B}_p^2}{\partial n}\nonumber \\
& -\frac{1}{8\pi\rho (r\sin\theta)^2}\frac{\partial(r\sin\theta B_\phi)^2}{\partial n} + \frac{v_\phi^2}{r}\frac{\cos\Theta}{\sin\theta}. \label{eqn_forcen_red}
\end{align}
Once the jet begins to collimate, the curvature term $-v_{\rm p}^2\partial \Theta/\partial s$ also becomes negligible. The magnetic force may act to either collimate or decollimate the jet, depending on the relative magnitudes of the toroidal $(r\sin\theta B_\phi)^2$ gradient (which collimates the outflow) and poloidal ${\mathbf B}_p^2$ gradient (which ``decollimates''); in our simulations, the collimation of the matter implies that the magnetic hoop stress is larger than the poloidal field gradient. Thus the main perpendicular forces acting in the jet are the collimating effect of the toroidal magnetic field and the decollimating effect of the centrifugal force.

\subsubsection{Analysis of the forces}
\begin{figure*}
\centering
\includegraphics[width=170mm]{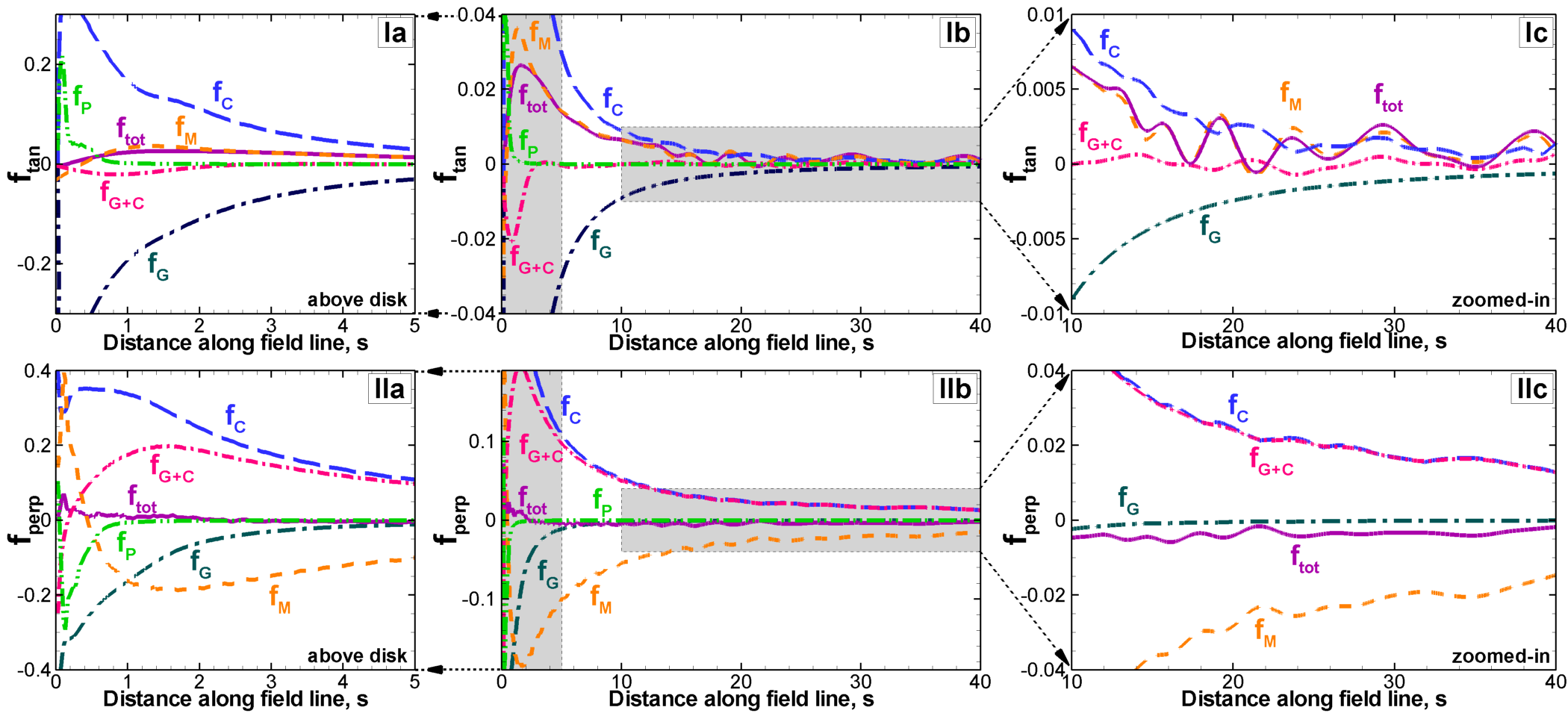}
\caption{Force components along a field line in the jet. The top row of {\bf panels (Ia), (Ib) \& (Ic)} show the {\it tangential} projection of the forces along the representative field line in: (Ia) the region just above the disk; (Ib) the whole simulation region;  (Ic) the region far above the disk. A positive tangential force indicates acceleration away from the equatorial axis (i.e. the disk plane).
{\bf Panels (IIa), (IIb), \& (IIc)} show similar plots of the {\it perpendicular} projection of the forces along the representative field line. A positive normal force indicates acceleration away from the axisymmetry axis.
\label{fig_sforce}}
\end{figure*}
Here, we analyze the role that the magnetic, gravitational, centrifugal, and pressure forces play in launching and collimating the outflow. Fig. \ref{fig_streamline} shows vectors denoting the magnitude and direction of the various forces along a ``representative'' field line running through the jet: this field line is the same as the one used previously for analysis of the velocities in \S \ref{subsec_veloc} and is anchored in the disk at \trad\ = 1.24. Fig. \ref{fig_sforce} shows the tangential and perpendicular projections of the individual forces onto this field line.

\paragraph*{Launching mechanism} The top row of Fig. \ref{fig_sforce} shows several different views of the tangential projections of the forces onto the representative field line. In order to accelerate the matter and launch an outflow, the sum of the tangential forces in the jet $f_{\rm tot,\parallel}$ must be positive overall.  The tangential projections of the gravitational and centrifugal forces are nearly equal and opposite everywhere in the jet (i.e. $f_{\rm G+C,\parallel} \approx 0$) except near the star where gravity dominates and acts to hold the matter in the disk (e.g. Fig. \ref{fig_sforce}Ia). In contrast, the magnetic pressure force $f_{\rm M,\parallel}$ acts in opposition to gravity and serves to accelerate the matter out of the disk along the magnetic field lines. Fig. \ref{fig_sforce}Ib shows that the gas pressure force $f_{\rm P,\parallel}$ is small outside the disk and plays no role in launching the outflow. Since $f_{\rm G+C,\parallel}$ and $f_{\rm P,\parallel}$ are both negligible far from the star, the magnetic force dominates within the jet giving $f_{\rm tot,\parallel} \approx f_{\rm M,\parallel}$. 

The effect of the various forces can be seen in the three panels of Fig. \ref{fig_streamline}: $f_{\rm G+C}$ pulls the matter downward toward the disk (Fig. \ref{fig_streamline}b) while the the magnetic force $f_{\rm M}$ acts to drive matter out from the disk (Fig. \ref{fig_streamline}c). When these forces are summed together, the resulting force $f_{\rm tot}$ shown in Figs. \ref{fig_streamline}a and \ref{fig_sforce}Ib is positive overall; note that even far from the disk, $f_{\rm tot, \parallel} > 0$ and the matter is continually accelerated. Since the centrifugal force is completely canceled by gravity, the jet is driven by a purely magnetic force and hence the launching mechanism is purely magnetic as well. This mechanism is similar to the inner disk wind model discussed in \citet{lovelace1991} and observed in simulations of conical winds in R09. 

\paragraph*{Collimation mechanism} As discussed previously in \S \ref{subsubsec_forceeqns}, the collimation of the jet is a competition between the decollimating effect of the centrifugal force (see Fig. \ref{fig_streamline}b) and the collimating force of the magnetic hoop-stress (Fig. \ref{fig_streamline}c); in order to collimate the jet, the sum of these perpendicular forces must be negative overall. The bottom row of Fig. \ref{fig_sforce} shows the {\it normal} projection of the various forces onto the representative field line. Very close to the star, the perpendicular component of the centrifugal force is large and the net force is positive, pushing some of the matter into a ``spur'' (see Fig. \ref{fig_streamline}a). However past $\ts = 2.5$, the centrifugal force weakens as the matter moves away from the rotation axis, allowing the magnetic hoop-stress to collimate the matter into a jet. Fig. \ref{fig_streamline} shows this very clearly: the centrifugal force in Fig. \ref{fig_streamline}b acts opposite to the magnetic force in Fig. \ref{fig_streamline}c. When the two forces are summed (Fig. \ref{fig_streamline}a), the forces largely cancel and the residual points inward toward the axis, serving to collimate the outflow into a jet. Note that the scale of the vectors in Fig. \ref{fig_streamline}a is much smaller than the scale in in Figs. \ref{fig_streamline}b and \ref{fig_streamline}c: the collimation is a delicate balance between the magnetic and centrifugal forces in the jet. 

\begin{figure}
\centering
\includegraphics[width=70mm]{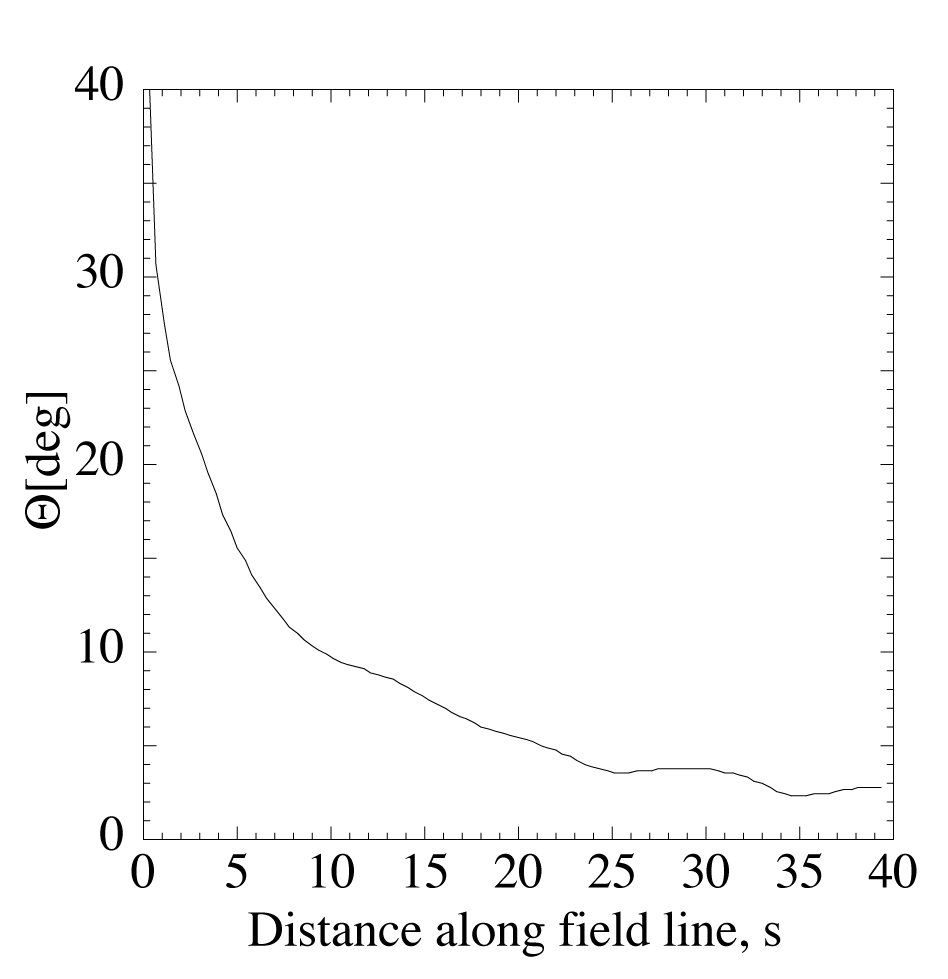}
\caption{Dependence of the angle $\Theta$ between
the reference poloidal field line and the axisymmetry axis
on the distance along the field line. \label{fig_slope}}
\end{figure}
 
Fig. \ref{fig_slope} shows the dependence of $\Theta$---the angle between the reference poloidal magnetic field line and the axisymmetry axis---on the distance along the field line \ts. For the poloidal field line in the reference case, we measure a half-opening angle $\Theta \approx 4^\circ$ for $\ts > 25$. 

\section{Discussion} \label{sec_discussion}
\subsection{Applications to young accreting stars} \label{subsec_applications}
The dimensionless form of the simulations permit us to apply the results to a variety of stars of different sizes, from neutron stars to young T Tauri stars. In R09, applications of the conical winds to different types of stars were discussed in detail and more recently, a model similar to the one presented in this paper has been applied to the rapidly accreting FU Orionis \citep{konigl2011}. In our simulations, the high \mdot\ disk compresses the magnetosphere almost to the surface of the star, and the model is well suited for understanding young stars undergoing periods of enhanced accretion such as FUORs, EXors and magnetized CTTSs.

\paragraph*{EXors} EXors represent an interesting stage in the evolution of young stars where the accretion rate may be enhanced up to $\mdot\sim10^{-6}-10^{-5} {\rm M}_\odot/{\rm yr}$ resulting in the ejection of powerful outflows \citep[e.g.][]{coffey2004, brittain2007}. The accretion rate obtained in our simulations is  $\mdot = \widetilde{\mdot} \mdot_0$  (where $\widetilde{\mdot}\approx 160$ is the dimensionless accretion rate obtained in simulations, see Fig. \ref{fig_flux}). The reference accretion rate $\mdot_0$ (and the dimensional accretion rate) strongly depend on the magnetic field of the star. For EXors, we choose the value  $B_*=800$ G  and obtain  $\mdot\sim1.1\times 10^{-5} \msun/{\rm yr}$. Table \ref{tbl_refval} shows the other reference values for this case.

\paragraph*{FUORs} The model is also well suited for describing FUORs for which the accretion rate is very high, $\mdot\sim10^{-5}-2\times10^{-4} \msun/{\rm yr}$, and the disk can strongly compress the magnetosphere. In  \citet{konigl2011} the properties of the wind (density, velocities) were compared with those derived from the blue-shifted spectral features observed in the wind of FU Ori \citep{calvet1993, hartmann1995} and reasonable agreement was found. Table \ref{tbl_refval} shows an example of the reference values for FUORs. Here, we increased the magnetic field of the star up to $B_*=2.5\times10^3$ G in order for the magnetosphere to be able to stop the disk near the star. For this $B_*$, we obtain an accretion rate of $\mdot = \widetilde{\mdot} \mdot_0 = 2.5\times 10^{-4} \msun/{\rm yr}$ \footnote{Note that \citet{konigl2011} adopted a larger stellar radius and a smaller \mdot, finding $B_*=2.1$ kG.}.

\paragraph*{CTTS} The model can be also applied to young stars---such as protostars and CTTS---with much lower accretion rates. Compared to FUORs, CTTSs show relatively small accretion rates of $\mdot \approx 10^{-7}-10^{-8}\msun/{\rm yr}$. Typical magnetic field strengths around T Tauri stars are on the order of one to a few kilogauss \cite[see][]{yang2011}. However, the stellar field may be dominated by higher-order components of the magnetic field while the dipole field (which dominates in the disk) may be comparatively weak \citep{donati2007, gregory2008, donati2011}. Taking this into consideration, we choose a weak equatorial dipole field of $B_* = 100$ G as a CTTS reference value (see Table \ref{tbl_refval}) and obtain a disk accretion rate of $\mdot = \widetilde{\mdot} \mdot_0 = 1.8\times 10^{-7} {\rm M}_\odot/{\rm yr}$.

The ejection-to-accretion ratio obtained in our model, $\mdot_{\rm wind}\approx 0.2 \mdot_{\rm accr}$ (see Fig. \ref{fig_flux}), is somewhat larger than the typical values observed in CTTSs ($\lesssim$0.1) or FUORs ($\sim$0.1) \citep{hartmann1996, coffey2008}. In our simulations, the jets are successfully launched when the initial outflow is able to penetrate through the matter-dominated corona. The coronal density in our simulations is quite low (typically $\sim10^{-4}$ of $\trho_{\rm d}$). Despite this, the corona is initially matter-dominated because the dipole magnetic field strength decreases rapidly with distance. The many simulations which produced collimated jets (see Tab. \ref{tbl_runs}) showed that the magnetic launching mechanism is robust and can drive jets from disks with a wide range of densities and accretion rates. However, only the most powerful outflows can penetrate through the matter-dominated corona. It is possible that we observe a selection effect in which only most powerful jets successfully penetrate through the dense corona, and in reality (when the coronal density is even lower),  weaker jets can be launched. In that case, we expect the matter ejection-to-accretion ratio $\mdot_{\rm wind}/\mdot_{\rm accr}$ to be lower.

\subsection{Collimation in different simulations} \label{subsec_collimdiscuss}
In Tab. \ref{tbl_runs} we show a sample of the simulations which produced collimated jets. The simulations exhibited a wide range of degrees of collimation with half-opening angles ranging between $\Theta=4^\circ$ and $\Theta=20^\circ$. The collimation observed in the jets is stronger than in the conical winds studied in R09\footnote{We should note that even in the case of the conical winds, the half-opening angle at the boundary reaches $20^\circ-25^\circ$ at the end of the simulation run (see Fig. \ref{fig_timestep}, t = 700 panel) which is marginally close to the least collimated jets observed in current simulations.}.

The key difference between the simulations presented in R09 and the new simulations presented here is the size of the simulation region: the new simulation region is nearly triple the size of the regions in the
previous work. The larger simulation region in our model setup results in a higher accretion rate\footnote{At the external boundary we fix the density of the incoming matter and hence \mdot\ increases with the size of the simulation region.}. In addition, in the larger region, incoming disk matter collects and compresses a larger portion of the star's magnetic flux toward the star. Both factors lead to a higher magnetic pressure at the disk-magnetosphere boundary and to a correspondingly higher magnetic force. This results in the emergence of a more powerful, more magnetized outflow, leading to stronger collimation of the jet. We measure magnetization level with the kinetic plasma parameter $\beta_1$ which takes the ratio of the thermal plus ram pressure to the magnetic pressure (Eq. \ref{eqn_beta}). In all of the cases shown in Tab. \ref{tbl_runs}, the jets are matter-dominated (i.e. $\beta_{\rm 1, jet}>1$). However, in strongly matter-dominated jets ($\beta_{\rm 1, jet} \apprge 10$), weaker collimation is observed---for example, in the conical winds observed in R09, $\Theta\sim$30$^\circ$ and $\beta_{\rm 1, jet}=10-30$. In contrast, when the outflow is only slightly matter-dominated (like in the reference case of this paper where $\beta_{\rm 1, jet}\approx 1-6$, $\Theta\approx4^{\circ}$) then collimation is strong. In \S \ref{subsec_properties} we argue that the spurious toroidal field at $\theta < 3.6^\circ$ has no influence on the jet collimation.

Recently, \citet{kurosawa2011} performed a special set of simulations for a magnetized star with much smaller accretion rates (as in a CTTS) where the disk does not compress the magnetosphere as strongly and is truncated at a few stellar radii. These simulations are much more applicable to CTTSs which usually have large magnetospheres and showed that outflows are usually less collimated when the accretion rate is low.

\subsection{Comparison with observation} 
Recent observations of several CTTSs with outflows have shown that $\sim$10 AU away from the star, the outflows have already become collimated. However, in almost all of the cases the degree of collimation is low, with opening angles of $\sim$20--30$^\circ$ at distances 10--50 AU \citep[e.g.][]{hartigan2004}. These outflows eventually become very well collimated at larger distances, with opening angles of $2-5^\circ$ further out \citep{dougados2000}. The jets in our simulations exhibit much faster collimation---$4-20^\circ$ at $0.4$ AU and are likely most relevant to highly accreting magnetized stars such as FUORs or EXors where the outflow is expected to be launched locally from a highly compressed disk-magnetosphere boundary. However, observational coverage of jets around FUORs and EXors is still sparse and there exist few constraints on the sub-AU structure of outflows from these highly accreting stars. However, as mentioned previously in \S\ref{subsec_applications}, comparisons of the our model with spectrally-derived properties of the wind in FU Ori star show that the simulations do agree with observed outflow velocities \citep{konigl2011}.

\citet{alencar2005} performed a spectral analysis of the $H_\alpha$, $H_\beta$, and NaD emission in the magnetosphere and at the base of the outflow in the highly accreting CTTS RW Aur A \citep[\mdot\ $\sim 10^{-7.5}$ to $10^{-6}\ \msun$/yr,][]{hartigan1995, white2001}. Their results suggest that the high-velocity microjet in RW Aur A is likely launched as a ``narrow wind'' from a disk region between 2.2--5 \rstar around the star, consistent with the launching region observed in our simulations. Their spectral fits also show that the field lines at the base of the outflow are inclined at $\Theta = 30-40^\circ$, consistent with the jet launching angle in our models (see Fig \ref{fig_slope}). 

High velocity outflows launched from the inner disk (such as the collimated jets in our simulations) may explain the central component of jets with a layered ``onion-skin'' structure where the highest velocity, well collimated portion of the outflow flows close to the axis \citep{bacciotti2000}. If the disk is also threaded by an ordered magnetic field, then an extended disk wind may be responsible for the outer, less-collimated, low-velocity layers \citep{ouyed1997, ferreira2006, fendt2000}. However, the simulations show a broad range of collimation and it is possible that the more weakly collimated jets are related to these outer layers as well.

\section{Conclusions} \label{sec_conclusion}

Through 2.5D MHD simulations, we have achieved robust, fully collimated jets emerging from the disk-magnetosphere boundary of an accreting magnetized star. We observe purely magnetic launching and collimation mechanisms: the cold matter at the disk-magnetosphere boundary is driven upward by the magnetic pressure and collimated by the helically wound magnetic field lines extending up from the disk. Approximately one-fifth of the incoming disk matter exits the system through the jet; the jet also advects some angular momentum out of the disk but the majority is transported outward through the disk by viscous stress. The degree of collimation in the jet is likely connected to the level of magnetization within the jet: our simulations show that strongly matter dominated jets with $\beta_1 \apprge 10$ are less collimated than the weakly matter dominated outflows where $\beta_1 \apprge 1$. In the reference simulation presented in this work, the matter in the jet is only slightly matter-dominated ($\beta_1 \gtrsim 1$) resulting in a magnetic field that strongly collimates the jet to a half-opening angle of $\Theta\approx4^{\circ}$ at the top of the simulation region. Previous simulations by our group performed in a smaller simulation region showed the emergence of weakly collimated conical winds with a larger $\beta_1$ in the outflow (R09). The high levels of magnetization necessary to produce a strongly collimated jets may arise around stars such as EXors or FUORs where a high accretion rate strongly compresses the magnetosphere. 

Observations suggest that the jets from protostellar systems become collimated at distances less than 10 AU away from the star (the present resolution limit) \citep{hartigan2004, coffey2008}. Our simulations show that the jet is launched and collimated on scales which are on the order of tens of stellar radii. However, the jets presented here may represent just the well collimated, high-velocity core of the ``onion-skin'' type outflows observed around young stars. 

\section*{Acknowledgments}
The authors thank G.~V. Ustyugova and A.~V. Koldoba for the development of the code used in the simulations, A. K\"onigl for discussion of FU Ori type stars, and Hui Li for discussion of the current flow in jets. We also thank the anonymous referee for valuable comments. This research was supported in part by NSF grant AST-1008636 and by a NASA ATP grant NNX10AF63G; we thank NASA for use of the NASA High Performance Computing Facilities.

\bibliographystyle{mn2e}
\bibliography{mn-jour,jets}

\appendix
\section{Reference Units} \label{appen_units}
We take the reference mass $M_0$ to be the mass $M_*$ of the star.  The reference radius is taken to be the radius of the star, $R_0 = R_*$. The magnetic field is a dipole field and its value at the equator, $B_*$, is chosen such that the accretion disk (with typical accretion rates for young stars) can compress the magnetosphere almost to the stellar surface. The reference magnetic field is $B_0 = B_*/{\mu}$, where $\mu$ is the dimensionless magnetic moment. The reference velocity is $v_0 = (GM/R_0)^{1/2}$, the Keplerian velocity at the surface of the star. We measure time in units of $t_0 = 2\pi R_0/v_0$ (the Keplerian rotation period at $r = R_0$). The reference force per unit mass is $v_0^2/R_0$. The reference density is taken to be  $\rho_0 = B_0^{2}/v_0^{2}$. The reference temperature is $T_0 = P_0/({\cal R} \rho_0) = v_0^{2}/{\cal R}$, where ${\cal R}$ is the gas constant. The reference mass accretion rate is $\mdot_0 = \rho_0 v_0 R_0^{2}$. The reference energy flux is $\dot E_0 = \mdot_0 v_0^{2}$. The reference angular momentum flux is $\dot L_0 = \dot M_0 v_0 R_0$. The reference current is $I_0 =c  R_0 B_0$. Tab. \ref{tbl_refval} shows examples of reference values for several different types of young accreting stars.

\begin{table}
\centering
\begin{tabular}{llll}
\hline \hline
          & CTTS    & EXors        & FU Ori    \\ \hline
{\bf initial} & & &				\\ \hline
\mstar [\msun]		&      0.8 &      0.8 &      0.5 \\
\rstar\ [\rsun]	&        2 &        2 &      2.5 \\
$t_*$ [days]		&     1.90 &     1.90 &     3.37 \\
$B_*$ [G]			&      100 &     800 &     2500 \\ \hline
{\bf derived} & & & \\ \hline
$R_0$ [cm] 			& $1.39\times 10^{11}$ & $1.39\times 10^{11}$ & $1.74\times 10^{11}$ \\
$t_0$ [days] 		&    0.366 &    0.366 &    0.648 \\
$v_0$ [cm s$^{-1}$] &  $2.76 \times 10^{7}$ &  $2.76\times 10^{7}$ &  $1.95\times 10^{7}$ \\
$B_0$ [G] 			&       10 &      100 &      250 \\
$f_0$ [dynes/gm]  &	$5.48 \times 10^{3}$ &	$5.48 \times 10^{3}$ & $2.19 \times 10^{3}$ \\
$I_0$ [G cm$^2$/s] &	$2.09 \times 10^{22}$ & $1.67 \times 10^{23}$ & $6.52 \times 10^{23}$ \\
$I_0$ [A]		&	$6.96 \times 10^{12}$ & $5.57 \times 10^{13}$ & $2.18 \times 10^{14}$ \\
$\rho_0$ [g cm$^{-3}$] & $1.31 \times 10^{-13}$ & $8.39 \times 10^{-11}$ & $1.64 \times 10^{-10}$ \\
$n_0$ [cm$^{-3}$] & $7.83 \times 10^{10}$ & $5.01 \times 10^{12}$ & $9.79\times 10^{13}$ \\
$\mdot_0$ [\msun\ yr$^{-1}$] & $1.11 \times 10^{-9}$ & $7.12 \times 10^{-8}$ & $1.54 \times 10^{-6}$ \\
$\dot E_0$ [erg s$^{-1}$] & $5.35\times 10^{31}$ & $3.43\times 10^{33}$ & $3.70\times 10^{34}$ \\
$\ldot_0$ [erg s$^{-1}$] & $2.70\times 10^{35}$ & $1.73\times 10^{37}$ & $3.29\times 10^{38}$ \\
$T_{d}$ [K] 		&     9178 &     9178 &      4589 \\
$T_{c}$ [K] 		&  $9.18\times 10^{6}$ &  $9.18\times 10^{6}$ &  $4.59\times 10^{6}$ \\ \hline \hline
\end{tabular}
\caption{Calculated reference values for various types of young, accreting stars. We choose the values of the stellar mass \mstar, radius \rstar, period $t_*$, and equatorial magnetic field $B_*$ of the star and derive the other reference values from these initial values. The stellar rotation period is set by the corotation radius of the disk which is set to $R_{\rm cor} = 3$. The magnetic field $B_*$ is chosen such that the disk stops close to the stellar surface. To apply the simulation results to a particular class of stars, one needs to multiply the dimensionless values by the reference values shown in this table. \label{tbl_refval}} 
\end{table}

\section{Viscosity and diffusivity in the disk} \label{subsec_viscosity}
The momentum flux-density tensor \tensor\ in the MHD equations (Eq. \ref{eqn_mhdmom}) can be expanded as
\begin{equation}
\tensor_{ik} = \rho v_i v_k + P \delta_{ik} + \left(\frac{B^2}{8\pi}\delta_{ik} - \frac{B_i B_k}{4 \pi}\right) + \tau_{ik}.
\end{equation}
Here, $P$ is the gas pressure, $\delta_{ik}$ is the Kronecker delta function, and $\tau_{ik}$ describes the viscous stress from small-scale turbulence in the velocity and magnetic fields. This non-ideal portion of the stress tensor, $\tau_{ik}$, is dominated by two components
\begin{equation}
\tau_{r \phi} = -\nu_t \rho r \sin \theta \frac{\partial \Omega}{\partial r}, \quad
\tau_{\theta \phi} = -\nu_t \rho \sin\theta \frac{\partial \Omega}{\partial r}, \label{eqn_visc}
\end{equation}
where $\Omega$ is the angular velocity and $\nu_t$ is the coefficient of the kinematic turbulent viscosity. In our simulations, we do not consider the effects of viscous heating or radiative cooling based on the expectation that the two processes will compensate for one another. Rather, the main role of viscosity is to facilitate the transport of angular momentum outward through the disk, thereby permitting matter to accrete inward toward the star.

To estimate the turbulent viscosity $\nu_t$, we adopt the \citet{shakura1973} $\alpha$ model which approximates the viscosity coefficient as $\nu_t = \alpha_v c_s^2 \xi / \Omega_K$, where $c_s = (P / \rho)^{1/2}$ is the isothermal sound speed, $\xi$ is a density threshold coefficient, and $\Omega_K$ is the Keplerian angular velocity. Similarly, we estimate the turbulent diffusivity coefficient in Eq. \ref{eqn_mhd} as $\eta_t = \alpha_d c_s^2 \xi / \Omega_K$ \citep{biznovatyi1976}. Both $\alpha_v$ and $\alpha_d$ are dimensionless coefficients which are treated as parameters in our model. $\xi$ is a coefficient which sets a density threshold for the viscosity and diffusivity 
\begin{equation}
\xi = 
\begin{cases}
0 & {\rm if}\  \trho \le \trho_{\rm d}/4 \\
\frac{\trho - \trho_{\rm d}/4}{\trho_{\rm d} - \trho_{\rm d}/4} & {\rm if}\ \trho_{\rm d}/4 < \trho\ < \trho_{\rm d} \\
1 & {\rm if}\  \trho \ge \trho_{\rm d} \\
\end{cases}
\end{equation}
where $\trho_{\rm d}$ is the disk density parameter at the external boundary (see Tab. \ref{tbl_runs}). The density threshold coefficient $\xi$ varies between 0 to 1 and acts to smoothly ``turn on'' viscosity and diffusivity for regions with $\trho\ > \trho_{\rm d}/4$. 
\begin{figure}
\centering
\includegraphics[width=85mm]{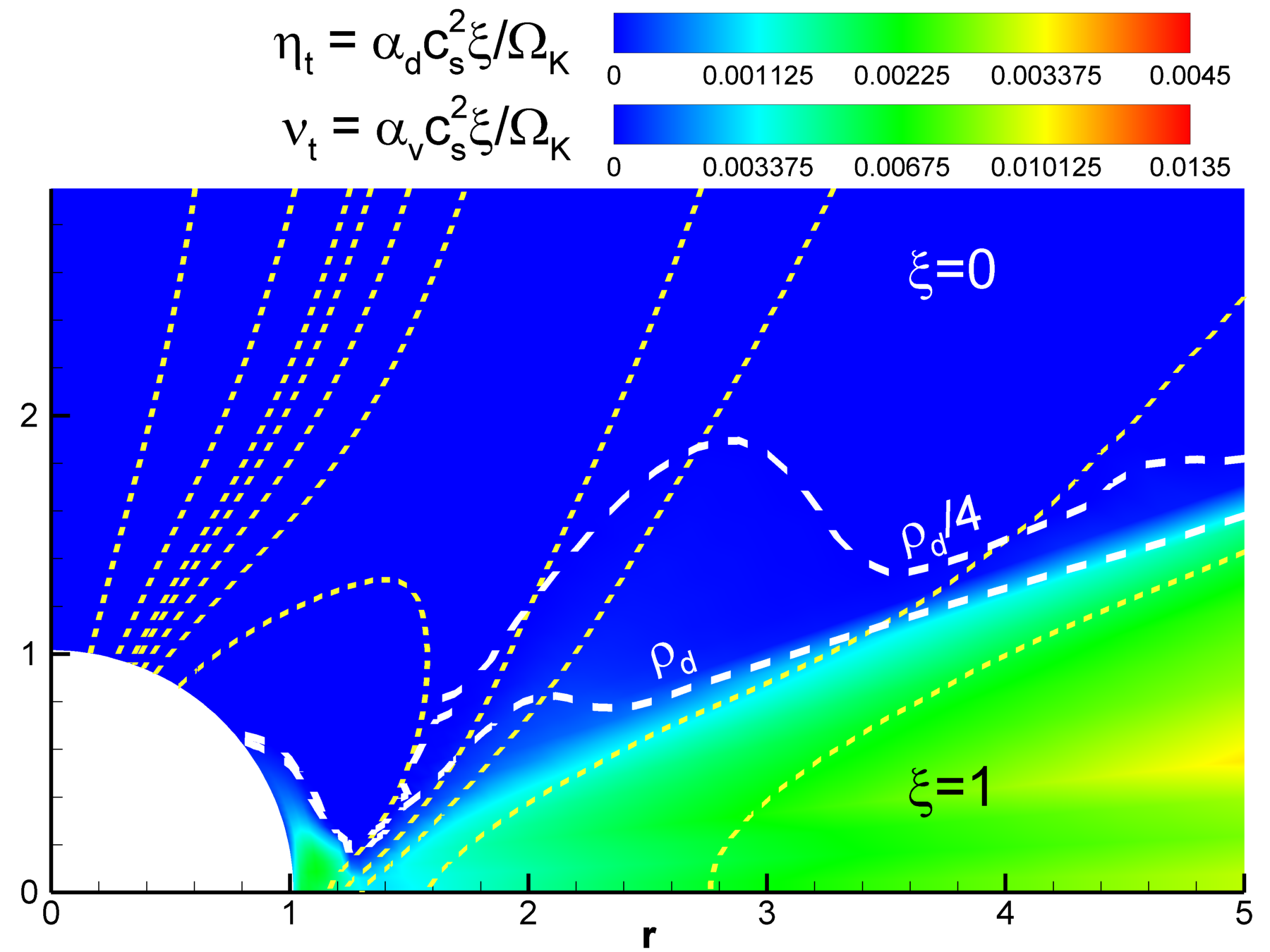}
\caption{Zoomed in plot of the diffusivity and viscosity profiles near the star at t = 860. In the $\xi = 0$ region, the turbulent viscosity $\nu_t$ and diffusivity $\eta_t$ are turned-off. Above the $\rho_d/4$ density threshold, both are smoothly turned on as $\xi \rightarrow\ 1$ and we solve the non-ideal MHD equations in the disk. \label{fig_diffusivity}}
\end{figure}
Fig. \ref{fig_diffusivity} shows the viscosity and diffusivity profiles of the inner disk of the reference simulation at \ttim = 860. Recall that the reference case used coefficents of $\alpha_d$ = 0.1 and $\alpha_v$ = 0.3 for the turbulent diffusivity and viscosity. At each timestep, the turbulent diffusivity and viscosity profiles are evolved and reapplied to every point on the grid to ensure self-consistency of the simulation. Above the $\rho_d/4$ ($\xi$ \textgreater\ 0) threshold, viscosity becomes important and we include $\eta_t$ and $\nu_t$ by numerically integrating the non-ideal MHD equations using an explicit conservative Godunov-type numerical scheme. Outside the disk where the density is low, $\xi = 0$ and we solve the equations of ideal MHD.

\section{Current flow in the simulation region} \label{subsec_current}
It is important understand the nature of the current flows in the simulation region. For general axisymmetric MHD flows the poloidal magnetic field ($[B_r, B_\theta]$ in spherical coordinates) can be written as
\begin{equation}
   {\bf B}_p =  {\bf \nabla} \times ( A_\phi \hat{{\rvecphi}}) = \frac{\hat{\bf r}}{r^2\sin\theta}
   \frac{\partial \Psi}{\partial \theta} - \frac{\hat{\rvectheta}}{r \sin \theta}
   \frac{\partial \Psi}{\partial r},
\end{equation}
where $\Psi(r,\theta) = r \sin\theta A_\phi(r,\theta)$ is the ``flux function'' for the poloidal magnetic field and $A_\phi$ is the vector potential \citep[e.g.,][]{lovelace1986}. The $\Psi ={\rm const}$ lines label the poloidal field in that ${\bf B}_p \cdot {\bf  \nabla }\Psi \equiv 0$.

For axisymmetric {\it and} non-relativistic MHD flows, Amp\`ere's law gives the
poloidal current density (in cgs units) as
\begin{equation}
   {\bf J}_p = \frac{c}{4 \pi} {\bf \nabla} \times ( B_\phi \hat{\mathbf \phi})=\frac{\hat{\bf r}}{r^2\sin\theta}
   \frac{\partial H}{\partial \theta} - \frac{\hat{\rvectheta}}{r \sin \theta}
   \frac{\partial H}{\partial r}\ ,
\end{equation}
where $H(r,\theta) = cr\sin\theta B_\phi(r,\theta)/4\pi$  acts as a the ``flux function'' for the poloidal current density  in that ${\bf J}_p \cdot {\bf \nabla} H \equiv 0$ \citep[e.g.,][]{lovelace1986}.

In cgs units, the total poloidal current through a spherical cap, $r={\rm const}$ and colatitude from  $0 ~{\rm to}~\theta$, is simply
\begin{equation}
I_{p, {\rm cgs}} = 2\pi \int_0^\theta  r^2 \sin\theta d\theta J_{pr} = 2\pi H(r,\theta)=
\frac{c}{2}r \sin\theta B_\phi(r,\theta),
\end{equation}
where $J_{pr}=\hat{\bf r}\cdot {\bf J}_p$. In all the plots we show the normalized current, $I_p = r \sin\theta B_\phi(r,\theta)$, which is proportional to $I_{p, {\rm cgs}}$.
For the  dipole-type field symmetry about the equatorial plane assumed in our simulations, $\Psi(r,\theta)=\Psi(r,\pi-\theta)$; that  is, $\Psi$ is an even function about the equatorial plane.
This requires that the toroidal magnetic field be an odd function, $B_\phi(r,\theta)=-B_\phi(r,\pi-\theta)$ so that $B_\phi(r,\pi/2)=0$ \citep{lovelace1987}. Thus we necessarily have
\begin{equation}
I_p(r, \pi/2) = 0.
\end{equation}
Thus, our MHD simulations guarantee that the net current flow through  the
upper (or lower) hemisphere is exactly zero. A current outflow in a jet with, for example, $\theta <1$ is exactly balanced by a current inflow in the region $\pi/2 -\theta <1$. However, in the general case of {\it no} symmetry about the equatorial plane, we may have $I_p(r, \pi/2) \neq 0$ \citep[see][]{lovelace2010}.

\begin{figure}
\centering
\includegraphics[width=85mm]{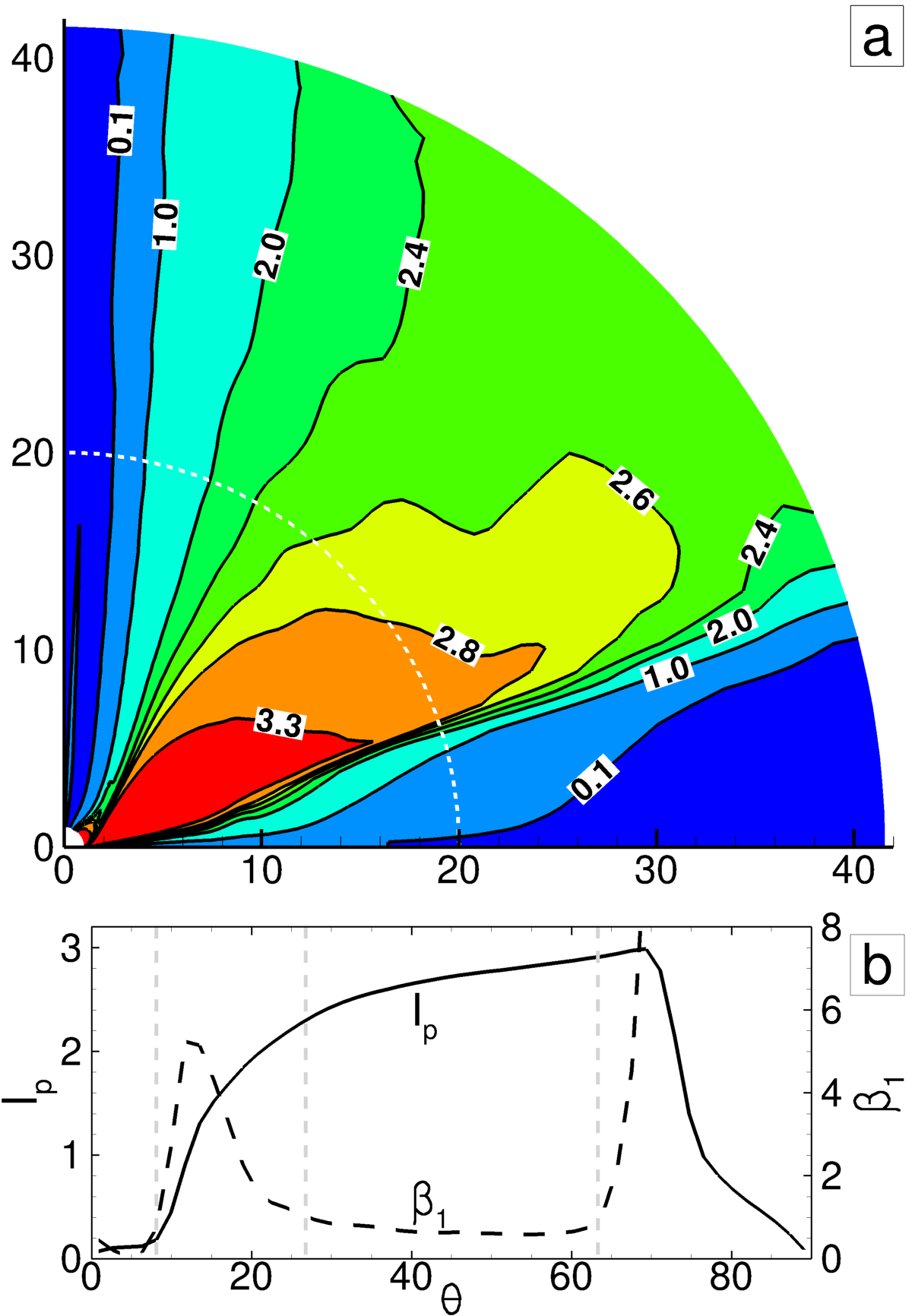}
\caption{\textbf{Panel (a)}:   The color background shows the values of $I_p(r,\theta)$ and the lines are current density lines. The numbers on the contours denote the local value of the current in normalized units. \textbf{Panel (b)}: The $\theta$ cross-section of the current and $\beta_1$ at $\trad\ = 20$, (see dashed line in the top panel). The dotted vertical lines represent the $\beta_1$ = 1 surfaces. \label{fig_current}}
\end{figure}

The top panel of Fig. \ref{fig_current} shows the poloidal current, $I_p(r,\theta)$, surfaces inside the simulation region. The values of $I_p$ in normalized units are shown on the lines.
   Fig. \ref{fig_current}b shows that the largest current flow is between the jet and the disk with a value $I_p = 3$, corresponding to $2.1\times 10^{13}$ Amp\`eres for a disk around a CTTS (see Tab. \ref{tbl_refval}). The current flow inside the jet is almost parallel to the jet, and is smaller than the current within the disk. The current carried by the jet flows out through the outer spherical boundary.  At the same time, an equal amount of  ``return'' current flows inward through the outer boundary and along the surface of the disk.  The jet current and  its return current have opposite signs and they repel as a result of their magnetic interaction mediated by the toroidal magnetic field $B_\phi$. This repulsion between the jet and its return current has been clearly observed and discussed in previous MHD simulations by \citet{ustyugova2000} and \citet{nakamura2008}. The toroidal magnetic field is responsible for collimating the jet as discussed in \S \ref{subsec_forces}.   At the same time the toroidal  field gives an outward radial force on the annular return current pushing it away from the jet axis.
\label{lastpage}

\end{document}